\documentclass{article}

\usepackage{arxiv}

\usepackage[utf8]{inputenc} 
\usepackage[T1]{fontenc}    
\usepackage{hyperref}       
\usepackage{url}            
\usepackage{booktabs}       
\usepackage{amsfonts}       
\usepackage{nicefrac}       
\usepackage{microtype}      
\usepackage{lipsum}
\usepackage{graphicx}
\usepackage{natbib}

\usepackage{booktabs} 
\usepackage{wasysym}
\usepackage{multirow} 
\usepackage{pdflscape}
\usepackage{threeparttable}
\usepackage{array}
\usepackage{tabularx}
\usepackage{enumitem}
\usepackage{caption}

\title{Going Down the Abstraction Stream with Augmented Reality and Tangible Robots: the Case of Vector Instruction}

\author{
Sergei Volodin \\
CHILI Lab \\
Ecole Polytechnique Fédérale de Lausanne \\
Lausanne, Switzerland \\
\texttt{sergei.volodin@epfl.ch} \\ 
\And
Hala Khodr \\
CHILI Lab \\
Ecole Polytechnique Fédérale de Lausanne \\
Lausanne, Switzerland \\
\texttt{hala.khodr@epfl.ch} \\ 
\And
Pierre Dillenbourg \\
CHILI Lab \\
Ecole Polytechnique Fédérale de Lausanne \\
Lausanne, Switzerland \\
\texttt{pierre.dillenbourg@epfl.ch} \\ 
\And
Wafa Johal \\
School of Computing \& Information Systems \\
University of Melbourne \\
Melbourne, Victoria, Australia \\
\texttt{wafa.johal@unimelb.edu.au} \\
}

\begin{document}
\maketitle
\begin{abstract}
Despite being used in many engineering and scientific areas such as physics and mathematics and often taught in high school, graphical vector addition turns out to be a topic prone to misconceptions in understanding even at university-level physics classes. To improve the learning experience and the resulting understanding of vectors, we propose to investigate how concreteness fading implemented with the use of augmented reality and tangible robots could help learners to build a strong representation of vector addition.

We design a gamified learning environment consisting of three concreteness fading stages and conduct an experiment with 30 participants. Our results  shows a positive learning gain. We analyze extensively the behavior of the participants to understand the usage of the technological tools -- augmented reality and tangible robots -- during the learning scenario. Finally, we discuss how the combination of these tools shows real advantages in implementing the concreteness fading paradigm.
Our work provides empirical insights into how users utilize concrete visualizations conveyed by a haptic-enabled robot and augmented reality in a learning scenario.

\end{abstract}


\begin{figure}[h]
    \centering
    \includegraphics[width=\linewidth]{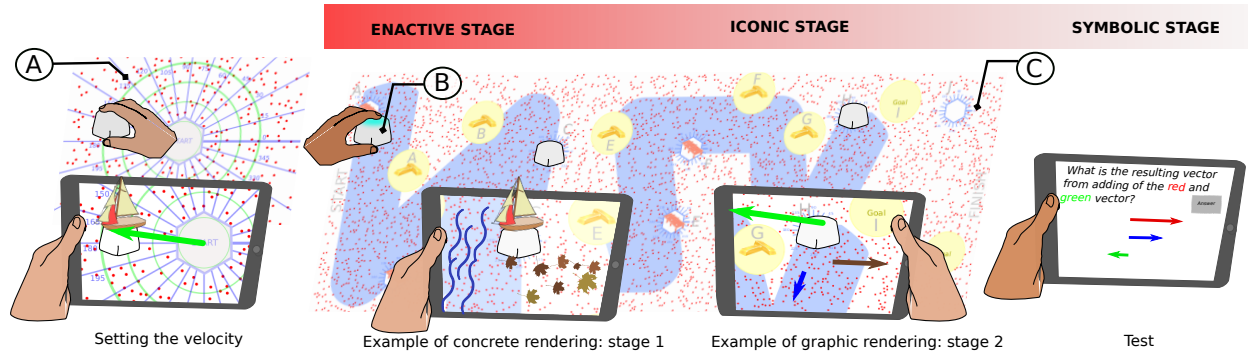}
    \caption{Activity Flow. The main working sheet (shown as C) contains 10 game levels located across a river (labeled A, B, ..., I) on the working sheet.
    For each of the levels, participants need to set the correct velocity to reach the gold (yellow circles with gold plates on the main working sheet).
    The velocity is set on a separate paper sheet (shown in A). The participant grabs the robot with their hand and moves it in the direction they want to set. The direction is shown in AR (below as a green arrow).
    Next, the robot is placed at the starting point (dock) of the level (shown in B), and the robot moves in the direction determined by the velocity, and also by the wind and the river's current. The tablet screens below the main sheet show the three stages of concreteness fading.}
    \label{fig:setup}
\end{figure}

\section{Introduction}

Graphical vector addition is studied in high-school and college, and it is a crucial component to understanding many scientific and engineering concepts, such as addition of velocities and forces in Physics, or geometrical computations in Computer Vision.
Classically, a vector is characterized by two components: a magnitude and a direction (including both sense/sign and orientation).
Despite being used in so many aspects of physics and mathematics, the studies show that it is difficult for college students to solve simple problems \citep{knight1995vector} related to vector addition.
Several studies showed that there are certain {\em misconceptions} or incorrect ideas that persist even after learning about vector addition \citep{nguyen2003initial,wutchana2011students}.
While investigating students' understanding of scalar multiplication of a vector, \cite{barniol2012students} identified several common mistakes, such as selecting an opposite direction, giving an incorrect magnitude or having issues with translating the vector to add them.
In addition, studying mathematics in general results sometimes in "mathematical anxiety" \citep{hembree1990nature,wiki:Mathematical_anxiety}, which could suggest that abstract concepts are difficult to learn, psychologically speaking. \cite{hembree1990nature} suggest that higher achievement can help students with mathematical anxiety.
We see the problems with studying vector addition as arising from mathematics being taught in a non-intuitive or too abstract way. It means that the rules of vector addition are not understood, and the concept is not grasped intuitively on simple examples \citep{nguyen2003initial, wutchana2011students}.
\cite{spiro1988multiple} have proposed to use multiple analogies to reduce the formation of domain-specific misconceptions and allow students to build more generic abstract models.
Following this, presenting students with concrete simple examples could be a fruitful way to introduce abstract concepts such as graphical vector addition.
This idea has been instantiated by the concreteness fading technique \citep{fyfe_concreteness_2014} which proposes to give students concrete examples before moving on gradually to more abstract representations.




Recent developments in augmented reality (AR) and robotics (i.e. cost reduction, robustness and reliability) allowed to introduce these technologies in multiple new settings into the classroom \citep{akccayir2017advantages,xia2018systematic}. One of the examples of robotics platforms designed for education purposes is the Cellulo robot \citep{ozgur2017cellulo}, a small versatile robot usable as a single low cost device with a haptic interface \citep{khodr2020allohaptic} or within a swarm \citep{ozgur2018declarative} of robots.
In this paper, we propose to explore the use of Cellulo and Augmented Reality as a way to instantiate the concreteness fading approach. Because the concept of vectors can be challenging for students to learn and that it is important in mathematics and physics, we chose to design our use case around teaching vector additions to our students. 

The learning activity we designed uses concreteness fading as a way to introduce vector addition, from intuitive and simple examples to abstract graphical representations.
In order to help with mathematical anxiety, the levels of our learning activity are designed to gradually increase in difficulty. In addition, we gamify the experience (such as, adding sounds for success and failure), and use toy-like 3D models inside the game, to make the learning less daunting.
Additionally, the game is played in teams, in order to distribute the problem-solving between team members and make it easier from a psychological point of view as well.
The principal goal of this research is to explore performances and usage of our technological tools (augmented reality and tangible robots) throughout the concreteness fading stages.


\section{Related Works}
\begin{figure}[ht]
	\centering
	\includegraphics[width=0.5\textwidth]{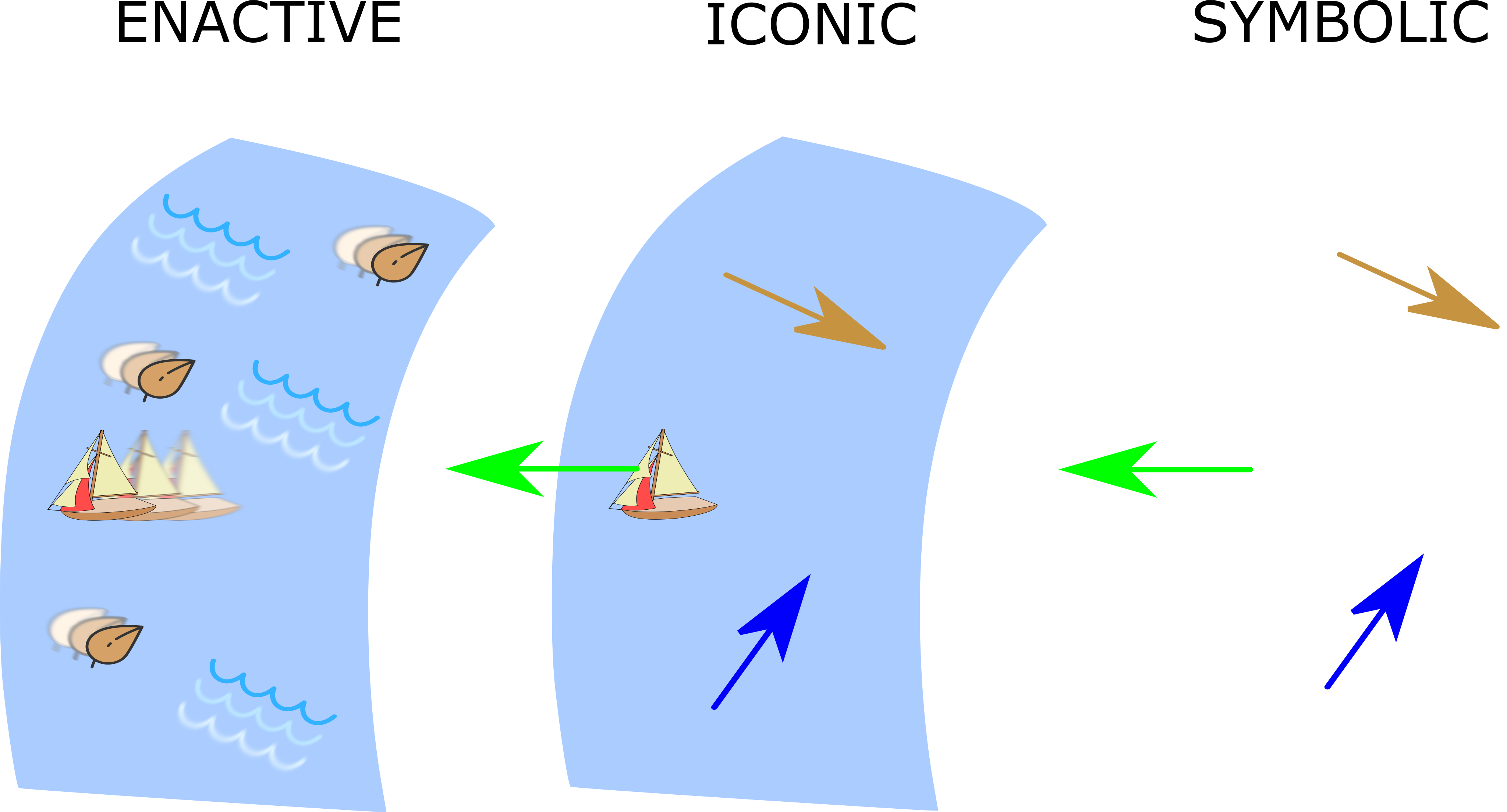}
	\caption{Illustration of concreteness fading. The physical/enactive representation refers to an actual moving ship. The pictorial/iconic representation refers to the notion of a moving ship, and the idealized/symbolic representation refers to the velocity vectors of the wind (brown), the river (blue) and the boat (green).}
	\label{fig:concreteness_fading}
\end{figure}


Piaget's theory of cognitive development explains that the different stages of learning involve various degrees of physical interaction, from sensorimotor stage in which children gain knowledge from physical interactions with the world, to formal operational stage which is characterised the ability to think and reason about abstract concepts \cite{piaget1970science}.
The use of concrete or abstract materials can both present benefits for learning.
Concrete materials are grounded in perception-motor experience, allowing students to easily identify links between form and its referent which they can connect to prior knowledge. 
Abstract materials are arbitrarily links to the referents: simpler, more transferable and generalizable across contexts.
\cite{fyfe_concreteness_2014} proposes to move beyond the "abstract vs concrete" debates and to investigate how these two types of material could co-exist temporally through \emph{concreteness fading}. 
\emph{Concreteness fading} refers to a three steps progression leading to instantiate a notion from concrete/physical to abstract over time \citep{Fyfe2019} (see \autoref{fig:concreteness_fading}):
\begin{enumerate}[noitemsep,nolistsep]
    \item Enactive: physical, concrete model of the concept
    \item Iconic: graphic, pictorial form of the concept
    \item Symbolic: abstract model of the concept
\end{enumerate}
Concreteness Fading is adopting Bruner's modes of representations: Concrete, Pictorial and Abstract. which argues that learning goes through the three stages to build a cognitive mental representation.
\cite{fyfe_concreteness_2014} found four main benefits of concreteness fading in mathematics and sciences: (1) it allows students to link abstract symbols with concrete objects, (2) it allows to ground abstract thinking with embodied and physical experiences, (3) it allows a multi-sensory memory storage of images that are associated to the abstract concepts, and (4) it allows for a generalization of the important properties.

Several researchers have provided successful examples of instantiating of the Concreteness Fading Theory in education \citep{fyfe_benefits_2015,ottmar2017concreteness,ching2019concreteness}. \cite{mcneil_concreteness_2012} propose a learning scenario in mathematics about operational rules (i.e. commutativity, associativity). Using a two step fading, authors demonstrate that using operands that go from concrete and meaningful operand (i.e. measuring cups) to abstract symbols (e.g. circles) allows students to obtain better transfer performance when compared to either concrete or abstract condition. While this research informs us in the combination of concrete and abstract, the material used was not as much embodied; and \cite{novack2014action} investigated exactly that by proposing a movement-based mathematical learning activity.


Finally, \cite{fyfe_benefits_2015} showed the benefit of the progression from concrete to abstract in a mathematical scenario, highlighting the importance of the fading process towards abstraction and generalization. 
One key point in using concrete material and concreteness fading is the link between the different representations.
\cite{brown2009using} warned practitioners on the difficulty to connect concrete to abstract representations (such as, the link between gestures and components of the concept should be smooth). \cite{Fyfe2019} also aimed to make the concreteness fading technique more practical by presenting examples and redefining the key concepts involved to get better adoption by teachers.
Building the material for smooth transition is indeed one of the main challenges for practitioners. While we found in the literature many examples instantiating of the theory using paper or gestures, there have been very few examples of the use of technology in this matter. 
The first work we find using digital tools in this context, by \cite{trory2018designing}, is using the concreteness fading approach to train students in a computational concept (router networking). Authors discuss the potential of a digitally augmented version of concreteness fading approach that could allow more gradual fading, one that could feature adaptive strategies to move toward abstract representations according to each child's pace.


In this study, we propose the implementation of the concreteness fading theory with a smooth three-step mode of representation (see Figure \ref{fig:cellulo} and Section \ref{sec:learning_scenario}) using augmented reality combined with tangible robots as a novel way to teach students about vector addition.
With this, we aim to investigate whether such an approach leads to a positive learning gain, and to analyze the behavior of the participants when moving between concreteness fading stages. We make the following hypothesis:
\begin{itemize}
    \item[H1:] Usability -  Robotics and AR technology can be used to design embodied to abstract experience of mathematical concepts such as vectors.
    \item[H2:] Learning Gain - Students can learn about vector addition  using our concreteness fading learning game.
    \item[H3:] Performance - Students performing well during our learning activity get a better learning gain. 
    \item[H4:] Students' behavior - We can identify interaction patterns in the use of AR and the robot that can predict positive learning gain.
\end{itemize}


\section{The Augmented Robotics Platform} 
\begin{figure*}[h]
    \centering
    \includegraphics[width=0.8\textwidth]{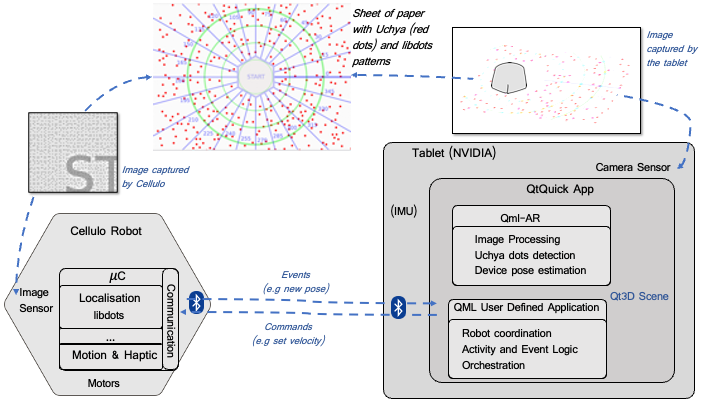}
    \caption{Software architecture using the qml-AR library and libdots \citep{Hostettler} for the tablet and robot localisation respectively}
    \label{fig:system}
\end{figure*}

The proposed system uses three main components (see Figure \ref{fig:setup}): a tablet -- used to display the information via Augmented Reality); the Cellulo robot -- to represent the pirate ship; and two sheets of paper -- one on which students can set the velocity of their ship and the other with the river and the different levels of the game. The game is played in teams\footnote{See \url{https://youtu.be/B4-2qYsAKH4} for a video overview of the game}.
Figure \ref{fig:system} illustrates the software architecture of the activity and details in particular the paper-based localisation system of the Cellulo robot and the tablet. 
Our goal with the learning system is to use the AR visualization to gradually transition from concrete representations (e.g. leafs animations and water flow) to abstract representations (e.g. vector arrow). We are using a tablet to allow the team members to see the AR scene. The scene is rendered on the workspace (defined by the robots on the paper sheets) by first determining the pose of the tablet camera relatively to the plane on which students are manipulating the robot.

While marker-based AR systems provide an absolute referential (unlike systems based on inertial measurement units \citep{Oskiper2012}) they are also more accurate than electromagnetic methods like FM and Bluetooth \citep{tsai2016application}, and easier to setup compared to Infrared \citep{pintaric2007affordable}.
The most popular QR-code-like markers have a disadvantage of being easy to spot, and can occlude parts of the working space. Seamless markers on the other hand look more like the background and thus are less noticeable. For our study, we chose to use Uchiya's \citep{uchiyama2011random} Random Dot markers that allows use randomly placed dot pattern to represent the marker.

We use Nvidia Android tablets with a Qt/QML application. For this experiment using AR, we developed a new library written in C++/QML called \emph{qml-ar} that optimizes the Uchiya library and provides a Qt3D/OpenGL integration.
\emph{qml-ar}\footnote{\emph{qml-ar} is available as an open source library \href{https://github.com/chili-epfl/qml-ar}{https://github.com/chili-epfl/qml-ar}} uses random dot markers \citep{uchiyama2011random}, and proposes a camera-based AR framework for developing QML applications. The library works with occluded (partially visible) markers and from various distances. The pre-processing is done on the GPU. The library has competitive performance on Android devices with 30 frames per second and 30 millisecond latency and works in ranges of approximately $0.1-2$ meters from the paper sheet with the dots.
The application is connected to the Cellulo robot, one per team (representing the pirate ship). 
Cellulo robots are graspable active tangibles which can also move on their own \citep{ozgur2017cellulo}.
The robots can localize themselves on paper thanks to a printed microdots pattern and a camera underneath the robots \citep{Hostettler}. The localization system outputs the absolute pose of the robot on the paper which is then sent to the connected tablet via Bluetooth.
The paper sheets defining the workspace have a secondary set of random Uchiya dot markers which allow for the augmented reality module to work.
The Cellulo robots were also used for their ability to render planar haptics feedback \citep{celluloTechIntro} which was previously used in various learning activities such as writing \citep{asselborn2018bringing,neto2020}, symmetry \citep{cellulo-sym}, linear functions \citep{khodr2020allohaptic} and computational thinking \citep{Nasir:270077}.


\section{Design of the Learning Scenario}
\label{sec:learning_scenario}

The addition of vectors is present at many stages of students' educational journey and often finds its practical application in Physics class. 
In this work, we propose a learning game in which students have to drive a pirate ship to collect a series of treasures located on the shores of a river (see Figure \ref{fig:setup}). 
In this context, students have to determine the correct initial velocity of their ship taking into account the current of the river, and the strength of the wind; i.e. add the initial velocity of the ship to the river's current to obtain the actual velocity of the ship on the river.
The game consists of a series of levels, each one is more abstract than the previous one. 
The whole activity passed through three stages of concreteness fading. Augmented Reality is used as a way to implement concreteness fading in a smooth and continuous way for the two first stages. We implemented it within three main states in the activity: intuitive gamified (enactive stage), graphical addition (iconic stage), and the post-test with abstract vectors (symbolic stage). Within the game, the smooth transition is made by overlapping elements from the previous level to the next.

Before the beginning of the activity, the experimenters asked students to form pairs. 
Each pair was given a tablet on which the application used throughout the experiment is running. 

\subsection{Introduction}
The experiment started with an introduction presenting on each of the tablets animations showing the goal of the game and how the river's direction alters the course of the ship. Next, it is suggested that the solution is to counteract the river when it is not going in the desired direction.

\subsection{Pre-test}
\label{sec:pre-test}
The pre-test consists of a series of 10 questions on vector addition. It asks to find a sum of, or the difference between two vectors graphically (see Test on Figure \ref{fig:setup}). 
The different test questions present various complexity of tasks, and the display of the questions varies as well. Specifically, the axes and ticks are visible at the beginning, but disappear at the later stages. For the first questions, all vectors start from the origin of the axes, and the start point of the answer vector is unchangeable. Towards the end of the test, the summand vectors start from non-origin points, and the answer vector's start point becomes variable as well.
The test is taken by the team via the tablet application.

After the pre-test, students are given a robot connected to their tablet application. The robot is representing the pirate ship. 
Each team is given an A3 setting map, (Figure \ref{fig:setup}  A )
used to set the velocity's direction of their ship.
This map features the Uchiya dots allowing to see the velocity vector in AR and printed graphics that indicates the angles similarly to a 360 degrees protractor.

\subsection{Activity Levels}
\label{sec:levels}
Students are assigned a river map fixed to the floor. Two teams would share the same river map. In order to avoid collisions of robots and too much overlap of the workspace, the two teams on each river map are given a different set of levels to solve such that: \emph{Stream 1} needs to reach treasures $A, C , E, G, I$ , and, \emph{Stream 2} needs to reach treasures $B, D, F, H, J$.  
Having the two teams co-located on the same map (with odd and even levels) aims to introduce competition and engagement (similar to a race). 
At each level, teams start at a dock and need to find the correct angle for their robot ship to reach the treasure of the same letter.
If they fail, their tablet would play a crashing sound when arriving on the shore and they would have the possibility to make their robot go back to the start by pressing a button on their tablet.
If they succeed, a winning sound would be played by their tablet and their robot ship moves automatically to their next dock.
The sounds played at each time a team loses or wins, also participates in the competition between teams.
Each level is timed with an 8-minute limit and students can try as many times as they want until they succeed or that the timer ends.
If the timer ends, their ship robot moves to their next dock.

\begin{figure}
	\centering
	\includegraphics[width=0.49\textwidth]{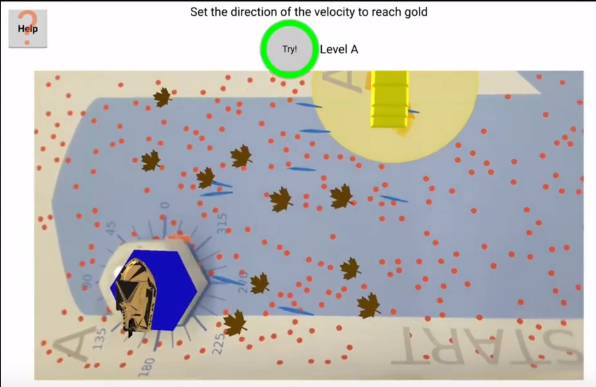}
	\hfill
	\includegraphics[width=0.49\textwidth]{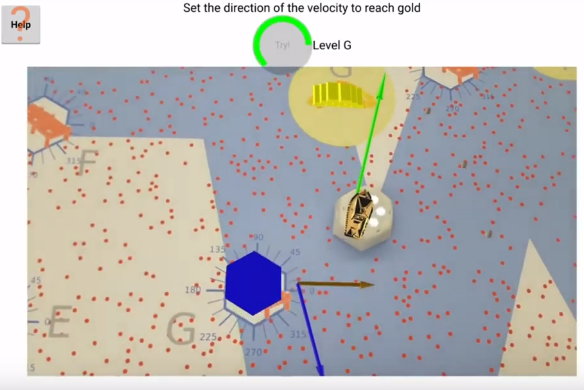}
	\caption{Top: screenshot of the level "A" in the game showing the ship (robot) at the dock (bottom left), leaves and waves showing the direction of the wind and the current. Bottom: screenshot of the level G in the game showing arrows (brown for the wind, and blue for the current, green for ship's velocity)}
	\label{fig:levels_A_G}
\end{figure}

\subsection{\textbf{Levels 1 (AC, BD): Enactive Stage}}
In the first group of levels (A\&C for Stream 1, and B\&D for Stream 2) augmented reality was used to show the wind's and the current's velocities as an graphical animation on top of the river. The wind is displayed as moving leafs; and the water current as moving waves.
At this stage, the main robot, when looked at through the AR application, features a 3D ship at its top (see \autoref{fig:levels_A_G}). No arrows are used at this stage.
The difficulties at the level A and B are designed to be relatively easy in order for students to get habituated to use the material.

\subsection{\textbf{Levels 2 (EG, FH): Enactive + Iconic Stage}}
In the second group (E then G for Stream 1 and F then H for Stream 2) arrows are displayed in AR to depict the wind and water current velocities (see \autoref{fig:levels_A_G}). In these levels, it is less obvious which direction is to be set. Since the game has a global time limit, and each level has its own time limit as well, it is not possible to try all the velocity directions to find the answer with "brute force". Therefore, only those teams can pass the level that can correctly identify the right direction using vector addition (or an intuitive understanding of it at the first levels).

\subsection{\textbf{Level 3 (I, J): Iconic Stage}}
\label{subsec:ij_iconic}
The third group is marked by an absence of gamification: the graphical animations of the river is no longer shown, and gold is replaced with a finish mark on the printed map. 
In order to aid adding vectors graphically which is required for a successful completion of these levels, a "virtual canvas" inside the AR scene is used. This is a virtual space where it is possible to move and change vectors so that it is easy to use the graphical vector addition rule to predict where the ship will go (see \autoref{fig:graphical_addition}). Just like in the previous levels the ship's direction is set using the setting map, and the vector is displayed as an arrow in AR. After that the students can use the \emph{Virtual Canvas} as a drawing space in order to predict where the ship will go. The students' input on the tablet consists in drawing the actual speed of ship after performing the addition between initial velocity, water current and wind vectors. If it is not going in a correct direction, they are allowed to make changes to the initial velocity setting. 




\subsection{Post-test: Symbolic Stage}
The post-test featured same 10 questions presented in the pre-test but horizontally mirrored. In addition, there are multiple-choice questions based on \citep{nguyen2003initial} aiming to evaluate the basic knowledge of vectors. 

\section{Experiment}

\subsection{Participants}
High-school students of Year 10 and Year 12 (aged M=15.9, SD=1) were invited to participate to the experiment, each not having previous knowledge of the subject of vectors and vector addition according to their Mathematics teacher (see Table \ref{tab:participants} to see the team compositions). 
The experiment took place in a separate room of the high school campus that usually hold practical STEM activities. 
Students were asked to form pairs at the beginning of the experiment. 
Fourteen pairs were formed (7 for each year-level). 

\begin{table}[!ht]
    \centering
    \begin{tabular}{|l|l|l|l|}
    \hline
        name & Year & Gender 1  & Gender 2  \\ \hline
        Team 01 & Y10 & Male  & Male  \\ 
        Team 02 & Y12 & Female  & Female  \\ 
        Team 03 & Y10 & Male  & Male  \\ 
        Team 04 & Y12 & Male & NA  \\ 
        Team 05 & Y10 & Female & Female \\ 
        Team 06 & Y12 & Male  & Male  \\ 
        Team 07 & Y10 & NA  & Female  \\ 
        Team 08 & Y12 & Male  & Male  \\ 
        Team 09 & Y12 & Male  & Female  \\ 
        Team 10 & Y12 & Female  & Female  \\ 
        Team 11 & Y10 & Male & Male  \\ 
        Team 12 & Y12 & Male  & Male  \\ 
        Team 13 & Y10 & Male  & Male  \\ 
        Team 14 & Y12 & Male  & Male  \\ 
        \hline
    \end{tabular}
    \caption{Composition of the student teams with their year, gender and dominant hand.}
     \label{tab:participants}
\end{table}

\subsection{Experimental setup}
\begin{figure}[htbp]
    \centering
    \includegraphics[width=0.6\textwidth]{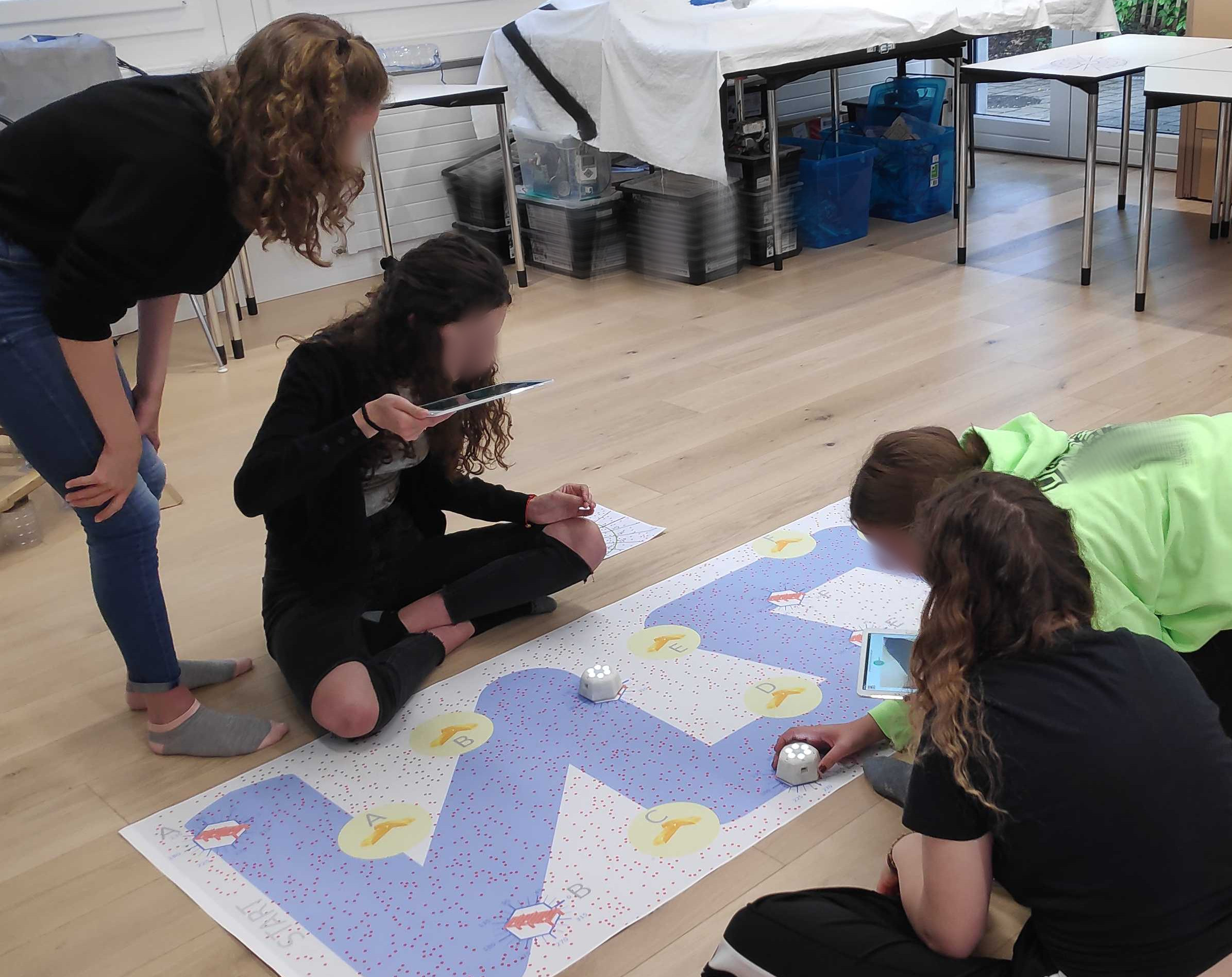}
    \centering
    \caption{Two teams using the AR on the tablet. The activity sheet on the floor contains two Cellulo robots}
    \label{fig:cellulo}
\end{figure}

The experiment was run in two sessions of 45min, one for the Year 10 and one for the Year 12 students (with the robot's activity lasting about 30min per session). Students were paired and given one robot (representing the ship), as well as a consumer-grade Android tablet for viewing the augmented reality scene, and the two paper sheets with the activity.
As shown on Figure \ref{fig:cellulo}, the large sheet of paper was placed on the floor to allow the students to navigate around the workspace freely and to enhance the AR performances. 
At a time, two pairs of students were working on the same river, but on different streams (see Section \ref{sec:levels}). Three experimenters were monitoring the activity and  solving technical issues (i.e. restarting the tablet app, connecting with the robots, etc).

\subsection{Experimental Metrics}
This experiment aimed to assess three main questions: 1) Do students learn about vector addition with our activity? 2) Is the performance during the game predicting students' learning gain? and 3) What type of interaction behaviors with the technology explains the learning gain? 



\subsubsection{Learning Evaluation Metrics}
To measure the efficacy of the activity from a learning perspective we consider 3 metrics:
1) Score on the pre-test, 2) Score on the post-test, 3) Score on a diagnostic quiz on vector concepts based on literature \citep{nguyen2003initial}.
The pre-and post-tests consisted on 10 questions asking teams to draw a vector on the tablet. These two tests were the same with the only difference being that one is horizontally mirrored version of the other. 
To compute the score for each question, we calculate the difference between the team's answer and the ground truth, in terms of angle and length of the vector. Since understanding the angle is the focus of our learning activity, we give a bigger weight to the angle than compared to the length ($10\colon 1$); making the score for each question a value between 0 and 1. All questions are weighted equally.
In addition to the drawing test, a diagnostic  multiple-choice questions (MCQ) test containing 4 questions on vector concepts understanding was administered to the teams. For this test, we report the weighted sum with the handcrafted weights for each question corresponding to the nature of the question (related to the direction or the length of the vector - since our activity mainly focused on teaching direction and not vector, questions on direction were given more weight). 

\subsubsection{Game Performance Metrics}
For each level, we consider the  direct direction as the one pointing from the starting dock to the gold. The naive direction does not result in a success for harder levels, because of the wind and the river's current. We measure the complexity of the level as the angular distance between the correct direction (leading to a success; reaching the gold) and the naive direction (vector pointing straight to the gold from the dock). Each level has an assigned concreteness fading stage (see Section \ref{sec:learning_scenario}), as well as the type (\emph{Stream 1} or \emph{Stream 2} depending if the team was starting at level A or B).

As metrics of game performance, we use 1) the normalized (with respect to average performance) number of attempts, as well as 2) the growth in number of attempts - a metric quantifying how well the teams learned during the game. For that, we treat for each team their normalized number of attempts as input points for a linear regression. After fitting the linear regression, we look at the slope of the line. A positive slope means that team takes more and more attempts to solve subsequent levels, compared to the average performance. A negative slope means that the team needs less and less attempts, and signifies learning in the team (relative to other teams). 

\subsubsection{Behavioral Metrics}
\begin{figure}
    \centering
	\includegraphics[width=0.25\textwidth]{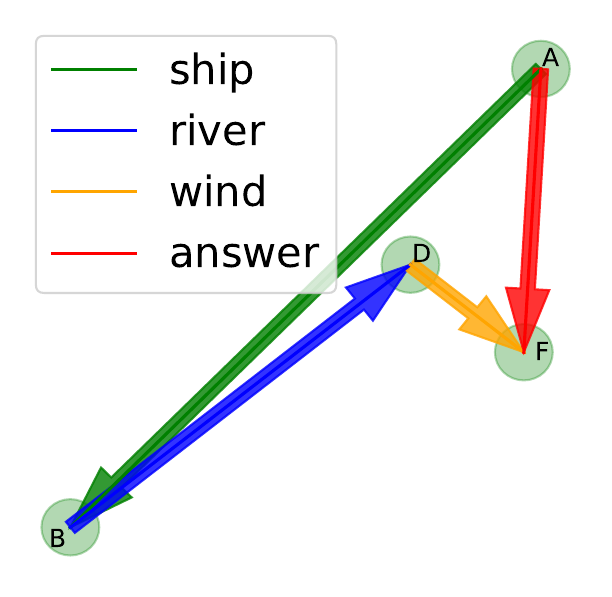}
	\caption{Graphical addition done right. To obtain the direction of the answer, or total ship's velocity correctly (red arrow), the participants need to add together the ship's relative velocity vector (green arrow) with the river's velocity vector (blue arrow), and the wind velocity vector (orange arrow). Green circles indicate that the adjacent vectors' ends are connected together. To do so, corresponding vectors' ends need to be moved moved to corresponding vector beginnings (they 'snap together' when close enough), and the answer vector needs to be set as the sum. The vectors are overlayed on top of the AR scene. Vectors can be added in any sequence.}
	\label{fig:graphical_addition}
\end{figure}

We define 3 main metrics as operationalizations of participants' behaviors during the learning activity. 
1) AR usage: 
AR was used as a medium to gradually transition from the enactive to the iconic stage.
We define two aspects for the AR usage. First, we calculate the average time the AR is used which we estimated by the number of times the AR visualisation was displayed for current game level (e.g. if the team was pointing the tablet correctly to the current working area and could see the AR). Second, we measure the jerkiness in the pose of the tablet, which is an indicator of how focused is the participant on AR. 2) Robot usage: we count the time the robot was located at the activity sheet, as well as the time it was grabbed/moved. This is computed using logs from the robot's localisation system and capacitive sensors present on top of the robot).
3) Completion of graphical addition level: We calculate if the graphical addition was used and identify if it was done correctly, for the last game level (Level I-J -- see Figure \ref{fig:graphical_addition}): one needs to add three vectors and set the total as the sum, each correct connection giving +1 to the total score for this metric (maximum 4 points).

\begin{figure}[ht]
	\centering
	\includegraphics[width=0.4\textwidth]{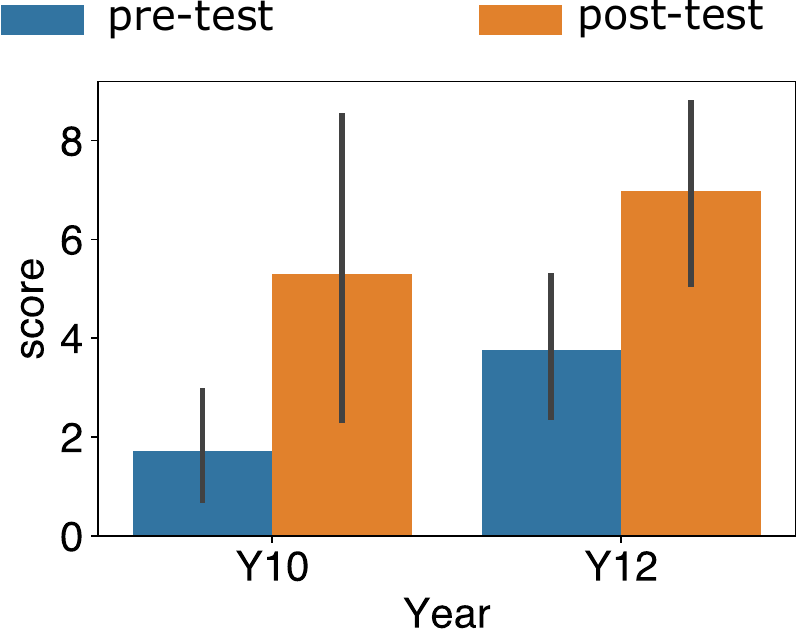}
	\caption{Pre- and post-test average score for Year 10 and Year 12 teams.}
	\label{fig:LG}
\end{figure}

\section{Results}

\subsection{Learning Gain and Tests}
For the pre-test, 43\% of questions are answered on average (SD = 20\%), and for the post-test, 72\% (SD = 26\%). The scores for the pre-test have a mean of 2.7 (SD = 2). For post-test, the mean score is 6.4 (SD=3). Since the data does not seem normally distributed, we used a Wilcoxon test to compare pre- and post-test scores. The test reveals that the pre- and post- test scores are significantly different ($T=10$, $p=5\cdot 10^{-3}$). The results show an increase of 50\% in correct answers between pre- and post-tests, therefore, giving evidence towards effectiveness of the game for learning vector addition.
Figure \ref{fig:LG} illustrates the Year-wise learning gain distribution computed from the pre- and post-test scores. The Kolmogorov-Smirnov test on pre- and post-test results does not show a difference between the two game streams -- starting with level A (Stream 1) or starting with level B (Stream 2), $p=0.64$ did not seem to influence the learning.

The MCQ test score has a significant OLS coefficient ($p=10^{-3}$) with post-test scores, meaning that scores at the MCQ are similar to the post-test scores.

The team's year class is correlated with the pre-test scores $\rho=0.62,\,p<10^{-4}$, but not with post-test scores. This shows that while there exists an initial difference between of knowledge between the two age groups (Year 10 and Year 12), it disappears after learning with our activity.

\begin{figure}[htb]
    \centering
    \includegraphics[width=0.4\textwidth]{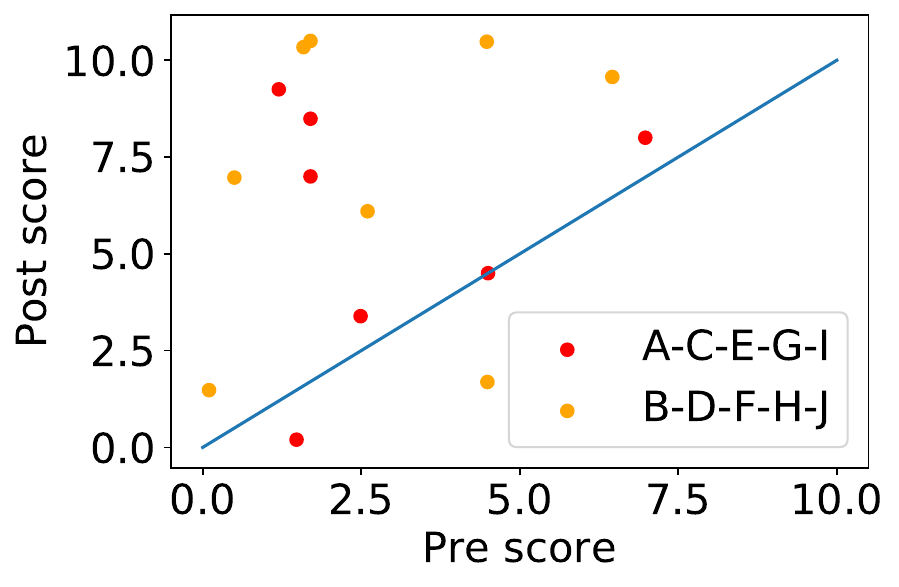}
    \caption{Distribution of pre- and post-test scores. The color indicates which levels were solved by the participants (A-C-E-G-I, Stream 1 or B-D-F-H-J, Stream 2)}
    \label{fig:prepost_xy}
\end{figure}

\subsection{Game performance}
\begin{figure}[htb]
\centering
\includegraphics[width=0.6\textwidth]{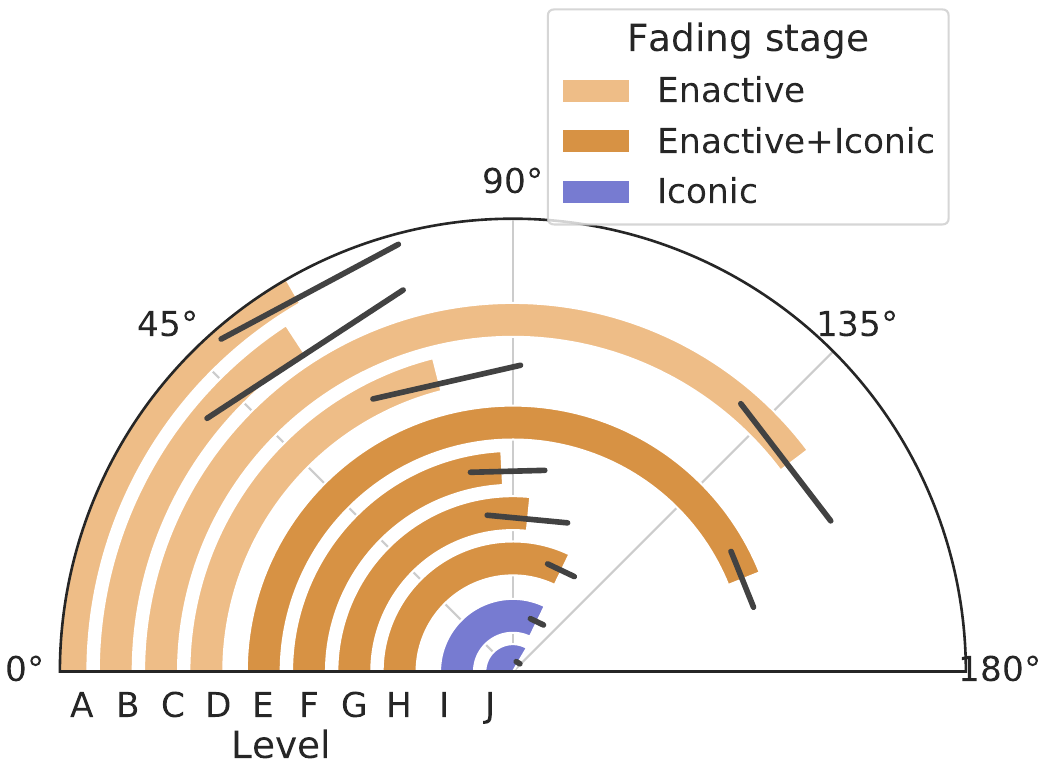}
\caption{Absolute Error Angle for Each Level (95\% confidence interval). The color of the bar indicates the concreteness fading stage. Errors are generally increasing as levels progress from concrete to abstract}
\label{fig:res_errorlevel}
\end{figure}

We divide each game into levels (A-C-E-G-I, Stream 1 and B-D-F-H-J, Stream 2), and treat each level solved by each team as a separate data point. This gives us 79 data points. For each attempt at each level, we verify the outcome manually (ship crashed/ship reached gold) to avoid software bugs.  Fig. \ref{fig:res_errorlevel} shows the average absolute error angle for each of the levels. 
Type of the game (Stream 1 or Stream 2) is negatively correlated ($\rho=-0.54, p<10^{-4}$) with the growth in number of attempts. 
While Stream 1 seems to induce more errors than Stream 2, the fact that there is no significant difference for the post-test scores (see previous section) can be explained by the fact that the game performance (mean number of attempts) loosely correlates with the post-test scores ($\rho=-0.40,\,p=3\times 10^{-4}$). In other words, while students perform differently in the two streams, they are able to adapt and achieve the same understanding of vectors regardless of the stream they trained on.
The class (Year 10 or Year 12) is correlated with the growth in the number of attempts ($\rho=-0.43,\,p=10^{-4}$). This shows that different age groups learn differently in the game (while having similar post-test scores). Younger participants struggle more with final levels and tend to try more.
The mean number of attempts is negatively correlated with the final post-test score ($\rho=-0.40,\,p=10^{-4}$). In other words, we see that less attempts result in higher scores. Since there is no correlation between pre-test scores and the number of attempts, this is not because of the initial difference in the teams' levels.

Per-class analysis shows a negative correlation ($\rho=-0.68,\,p<10^{-4}$) for Year 10 between the test gain score and the mean number of attempts, and a correlation of ($\rho=-0.89,\,p<10^{-4}$) between the growth in the number of attempts and the MCQ score. There is no significant correlation for the post-test score. A correlation involving the number of attempts does not exist for Year 12. For Year 12, the growth in the number of attempts is positively correlated with the learning gain ($\rho=0.76,\,p<10^{-4}$). This suggests that Year 10 students learnt more when making less attempts and fewer mistakes towards the end of the game, while Year 12 students learnt more when they make more mistakes towards the end of the game.

Growth in number of attempts is negatively correlated with the MCQ score $\rho=-0.48,\,p<10^{-4}$, which means that  less try-and-error behavior implies better multiple-choice responses.

\subsection{Behavioral Analysis}


    
The average time spent looking at the level in AR is positively correlated with the growth in the number of attempts ($\rho=0.50,\,p<10^{-4}$). This might suggest that teams which were struggling  used the AR more. 

Overall, the learning gain is not significantly impacted either by the robot usage or by the AR usage. Visual inspection of the AR usage(using the logs and looking at the visible AR workspace) and robot logs confirms it and shows similar patterns across all teams (looking at the level, touching the robot to set the velocity).
The test gain score is significantly impacted by the Virtual Canvas task (see \autoref{subsec:ij_iconic}) score $p=0.037$ (via OLS) with a positive coefficient, which shows that the final iconic stage is a good predictor of knowledge of vectors as assessed by the test.

We analyze the behavioral metrics, such as the jerkiness in the pose of the tablet. We obtain a result via causal analysis showing that more jerkiness in the angles results in better post-test scores (X="mean attempts", M="pose jerkiness", Y="post-test score", $X\to M\to Y$, $p<10^{-6}$). This is true for the jerkiness in the pose when holding over the velocity sheet, but not true for the main activity sheet. We hypothesize that it is because teams which are more confident in their knowledge of vectors are less stressed, and hold the tablet loosely when setting the velocity, making the tablet less stable.

\section{Discussion}
Learning through "tangible" examples is a well-established educational approach that helps students grasp complex concepts by starting with concrete, hands-on experiences before transitioning to more abstract ideas. The concretness fading paradigm formalises this approach proposing to smoothly transition from concrete to abstract concepts \cite{suh2020we}. 
In this paper, we proposed to operationalized concreteness fading with two types of technology: Haptic-robots and AR. 
We designed a learning game using the concreteness fading approach to introduce Year 10 and Year 12 students to the concept of vector addition. We used a tangible robots combined with augmented reality (AR) visuals to propose a smooth transition from concrete to abstract concepts when learning about vectors (H1 and H2 not rejected). 

One key finding of our study is the fact that our concreteness fading implementation allowed to smooth out the differences between students of Year 10 and Year 12 in terms of learning skills. 
Indeed, the transition from concrete to abstract representations of vectors allowed students from different age group to grasp the concepts equally well. This aspect of differentiation is crucial in educational context \cite{deunk2018effective}. 

We did not find significant differences in the way the teams were using the AR and robots to solve the tasks (e.g. in-task performance). This shows that the tools are accessible and well understood by students. 
However, unlike the usage of the haptic robot, the usage of AR did play an impact in leading to higher post-experiment learning gain for teams that were less relying on AR (H3 and H4 partially rejected). 
This can be explained by the fact that in the concreteness fading paradigm, embodied experience are important in the early stage of learning but towards the end of abstraction process, students are expected to rely less on their sensory perceptions but more on the mental model they have of the concept (e.g. its abstraction) \cite{fyfe_concreteness_2014}. 



Overall, the results show\begin{itemize}[noitemsep, nolistsep]
\item The \textbf{effectiveness of the activity}, via significant learning gains. While the initial levels of the two classes differed, they became similar after playing our learning game (H1 and H2). The activity was relatively short (30 minutes) compared to the time a teacher typically spends explaining vector addition concepts. Obtaining a learning gain and assessing transfer (from the activity to a typical test) is an important achievement. Evaluating student retention could ensure that the concepts were deeply learned.  
\item Only the performance at the last abstract level is correlated to learning (i.e., significant OLS score). This shows that students who successfully transitioned to abstraction during the activity performed the best on the post-experiment test (H3).
\item The game \textbf{difficulty levels are useful for learning} (i.e. correlation between number of attempts and both the post-test score and the multiple-choice test score) (H2) 
\item \textbf{Teams that learn better tend to use AR less}, demonstrating a successful abstraction of the concept (H4). This doesn't mean that AR was not useful but simply shows that students tended to rely less and less on the AR when advancing in the activity and it is a sign of abstraction.
\end{itemize}

Teams that learn better tend to use AR less, demonstrating a successful abstraction of the concept (H4). This does not mean that AR was not useful but indicates that students relied less on AR as they advanced in the activity, a sign of successful abstraction.

The goal of the study was not to compare binary modalities (AR vs. no AR, or Robot vs. no Robot) but to investigate the use of the technologies in a tailored activity that utilized both AR and tangible haptic robots. Our results showed that decreased use of AR towards the end of the activity correlated with higher learning gains, but robot usage was not a significant factor. The activity was designed based on recent research highlighting the advantages of robots in promoting embodied learning \cite{ioannou2020technology,tatarian2020mobiaxis}.

We also observed that teammates often naturally defined roles, with one student handling the tablet (i.e., viewing the AR) and the other using the robots to set the target direction. This division of resources is common in collaborative learning setups and can be beneficial to collaboration success, as it allows richer discussions from learners with different experiences and perspectives \cite{dillenbourg1995mediating}. Since the results reported in this experiment are at the group level, further work should investigate the collaborative mechanisms and how shared resources were managed within the teams.




\section{Conclusion}

In this paper, we proposed a new technological setup to implement the concreteness fading paradigm using tangible haptic-robots and AR. 
We assessed the usability, the educational benefit, the performance and the behavior of 28 students in a learning activity around vector addition. 
Our results provide support for the utility of AR technologies for learning, with no downsides
observed to the use of AR and tangible robots as a medium for implementing concreteness fading. Moreover, our results demonstrate that our implementation allowed to achieve significant learning gain for two different age groups. 
With this work, we foresee opportunities for future research to study how augmented robotics can be a medium for embodied collaborative learning.

\section*{Acknowledgement}
We would like to thank the teachers and the STEM Centre of the Ecole Internationale de Genève for their valuable support and collaboration.

\bibliographystyle{ACM-Reference-Format}
\bibliography{main}


\begin{thebibliography}{37}


\ifx \showCODEN    \undefined \def \showCODEN     #1{\unskip}     \fi
\ifx \showDOI      \undefined \def \showDOI       #1{#1}\fi
\ifx \showISBNx    \undefined \def \showISBNx     #1{\unskip}     \fi
\ifx \showISBNxiii \undefined \def \showISBNxiii  #1{\unskip}     \fi
\ifx \showISSN     \undefined \def \showISSN      #1{\unskip}     \fi
\ifx \showLCCN     \undefined \def \showLCCN      #1{\unskip}     \fi
\ifx \shownote     \undefined \def \shownote      #1{#1}          \fi
\ifx \showarticletitle \undefined \def \showarticletitle #1{#1}   \fi
\ifx \showURL      \undefined \def \showURL       {\relax}        \fi
\providecommand\bibfield[2]{#2}
\providecommand\bibinfo[2]{#2}
\providecommand\natexlab[1]{#1}
\providecommand\showeprint[2][]{arXiv:#2}

\bibitem[Ak{\c{c}}ay{\i}r and Ak{\c{c}}ay{\i}r(2017)]%
        {akccayir2017advantages}
\bibfield{author}{\bibinfo{person}{Murat Ak{\c{c}}ay{\i}r} {and} \bibinfo{person}{G{\"o}k{\c{c}}e Ak{\c{c}}ay{\i}r}.} \bibinfo{year}{2017}\natexlab{}.
\newblock \showarticletitle{Advantages and challenges associated with augmented reality for education: A systematic review of the literature}.
\newblock \bibinfo{journal}{\emph{Educational Research Review}}  \bibinfo{volume}{20} (\bibinfo{year}{2017}), \bibinfo{pages}{1--11}.
\newblock


\bibitem[Asselborn et~al\mbox{.}(2018)]%
        {asselborn2018bringing}
\bibfield{author}{\bibinfo{person}{Thibault Asselborn}, \bibinfo{person}{Arzu Guneysu}, \bibinfo{person}{Khalil Mrini}, \bibinfo{person}{Elmira Yadollahi}, \bibinfo{person}{Ayberk Ozgur}, \bibinfo{person}{Wafa Johal}, {and} \bibinfo{person}{Pierre Dillenbourg}.} \bibinfo{year}{2018}\natexlab{}.
\newblock \showarticletitle{Bringing letters to life: handwriting with haptic-enabled tangible robots}. In \bibinfo{booktitle}{\emph{Proceedings of the 17th ACM Conference on Interaction Design and Children}}. ACM, \bibinfo{pages}{219--230}.
\newblock


\bibitem[Barniol and Zavala(2012)]%
        {barniol2012students}
\bibfield{author}{\bibinfo{person}{Pablo Barniol} {and} \bibinfo{person}{Genaro Zavala}.} \bibinfo{year}{2012}\natexlab{}.
\newblock \showarticletitle{Students' difficulties with unit vectors and scalar multiplication of a vector}. In \bibinfo{booktitle}{\emph{AIP Conference Proceedings}}, Vol.~\bibinfo{volume}{1413}. AIP, \bibinfo{pages}{115--118}.
\newblock


\bibitem[Brown et~al\mbox{.}(2009)]%
        {brown2009using}
\bibfield{author}{\bibinfo{person}{Megan~C Brown}, \bibinfo{person}{Nicole~M McNeil}, {and} \bibinfo{person}{Arthur~M Glenberg}.} \bibinfo{year}{2009}\natexlab{}.
\newblock \showarticletitle{Using concreteness in education: Real problems, potential solutions}.
\newblock \bibinfo{journal}{\emph{Child Development Perspectives}} \bibinfo{volume}{3}, \bibinfo{number}{3} (\bibinfo{year}{2009}), \bibinfo{pages}{160--164}.
\newblock


\bibitem[Ching and Wu(2019)]%
        {ching2019concreteness}
\bibfield{author}{\bibinfo{person}{Boby Ho-Hong Ching} {and} \bibinfo{person}{Xiaohan Wu}.} \bibinfo{year}{2019}\natexlab{}.
\newblock \showarticletitle{Concreteness fading fosters children's understanding of the inversion concept in addition and subtraction}.
\newblock \bibinfo{journal}{\emph{Learning and Instruction}}  \bibinfo{volume}{61} (\bibinfo{year}{2019}), \bibinfo{pages}{148--159}.
\newblock


\bibitem[Deunk et~al\mbox{.}(2018)]%
        {deunk2018effective}
\bibfield{author}{\bibinfo{person}{Marjolein~I Deunk}, \bibinfo{person}{Annemieke~E Smale-Jacobse}, \bibinfo{person}{Hester de Boer}, \bibinfo{person}{Simone Doolaard}, {and} \bibinfo{person}{Roel~J Bosker}.} \bibinfo{year}{2018}\natexlab{}.
\newblock \showarticletitle{Effective differentiation practices: A systematic review and meta-analysis of studies on the cognitive effects of differentiation practices in primary education}.
\newblock \bibinfo{journal}{\emph{Educational Research Review}}  \bibinfo{volume}{24} (\bibinfo{year}{2018}), \bibinfo{pages}{31--54}.
\newblock


\bibitem[Dillenbourg and Schneider(1995)]%
        {dillenbourg1995mediating}
\bibfield{author}{\bibinfo{person}{Pierre Dillenbourg} {and} \bibinfo{person}{Daniel Schneider}.} \bibinfo{year}{1995}\natexlab{}.
\newblock \showarticletitle{Mediating the mechanisms which make collaborative learning sometimes effective}.
\newblock \bibinfo{journal}{\emph{International Journal of Educational Telecommunications}} \bibinfo{volume}{1}, \bibinfo{number}{2} (\bibinfo{year}{1995}), \bibinfo{pages}{131--146}.
\newblock


\bibitem[Fyfe et~al\mbox{.}(2015)]%
        {fyfe_benefits_2015}
\bibfield{author}{\bibinfo{person}{Emily~R. Fyfe}, \bibinfo{person}{Nicole~M. {McNeil}}, {and} \bibinfo{person}{Stephanie Borjas}.} \bibinfo{year}{2015}\natexlab{}.
\newblock \showarticletitle{Benefits of “concreteness fading” for children's mathematics understanding}.
\newblock   \bibinfo{volume}{35} (\bibinfo{year}{2015}), \bibinfo{pages}{104--120}.
\newblock
\showISSN{0959-4752}
\urldef\tempurl%
\url{https://doi.org/10.1016/j.learninstruc.2014.10.004}
\showDOI{\tempurl}


\bibitem[Fyfe et~al\mbox{.}(2014)]%
        {fyfe_concreteness_2014}
\bibfield{author}{\bibinfo{person}{Emily~R. Fyfe}, \bibinfo{person}{Nicole~M. {McNeil}}, \bibinfo{person}{Ji~Y. Son}, {and} \bibinfo{person}{Robert~L. Goldstone}.} \bibinfo{year}{2014}\natexlab{}.
\newblock \showarticletitle{Concreteness Fading in Mathematics and Science Instruction: a Systematic Review}.
\newblock  \bibinfo{volume}{26}, \bibinfo{number}{1} (\bibinfo{year}{2014}), \bibinfo{pages}{9--25}.
\newblock
\showISSN{1573-336X}
\urldef\tempurl%
\url{https://doi.org/10.1007/s10648-014-9249-3}
\showDOI{\tempurl}


\bibitem[Fyfe and Nathan(2019)]%
        {Fyfe2019}
\bibfield{author}{\bibinfo{person}{Emily~R. Fyfe} {and} \bibinfo{person}{Mitchell~J. Nathan}.} \bibinfo{year}{2019}\natexlab{}.
\newblock \showarticletitle{Making “concreteness fading” more concrete as a theory of instruction for promoting transfer}.
\newblock \bibinfo{journal}{\emph{Educational Review}} \bibinfo{volume}{71}, \bibinfo{number}{4} (\bibinfo{year}{2019}), \bibinfo{pages}{403--422}.
\newblock
\urldef\tempurl%
\url{https://doi.org/10.1080/00131911.2018.1424116}
\showDOI{\tempurl}
\showeprint{https://doi.org/10.1080/00131911.2018.1424116}


\bibitem[Hembree(1990)]%
        {hembree1990nature}
\bibfield{author}{\bibinfo{person}{Ray Hembree}.} \bibinfo{year}{1990}\natexlab{}.
\newblock \showarticletitle{The nature, effects, and relief of mathematics anxiety}.
\newblock \bibinfo{journal}{\emph{Journal for research in mathematics education}} \bibinfo{volume}{21}, \bibinfo{number}{1} (\bibinfo{year}{1990}), \bibinfo{pages}{33--46}.
\newblock


\bibitem[{Hostettler} et~al\mbox{.}(2016)]%
        {Hostettler}
\bibfield{author}{\bibinfo{person}{L. {Hostettler}}, \bibinfo{person}{A. {Özgür}}, \bibinfo{person}{S. {Lemaignan}}, \bibinfo{person}{P. {Dillenbourg}}, {and} \bibinfo{person}{F. {Mondada}}.} \bibinfo{year}{2016}\natexlab{}.
\newblock \showarticletitle{Real-time high-accuracy 2D localization with structured patterns}. In \bibinfo{booktitle}{\emph{2016 IEEE International Conference on Robotics and Automation (ICRA)}}. \bibinfo{pages}{4536--4543}.
\newblock
\showISSN{null}
\urldef\tempurl%
\url{https://doi.org/10.1109/ICRA.2016.7487653}
\showDOI{\tempurl}


\bibitem[Ioannou and Ioannou(2020)]%
        {ioannou2020technology}
\bibfield{author}{\bibinfo{person}{Marianna Ioannou} {and} \bibinfo{person}{Andri Ioannou}.} \bibinfo{year}{2020}\natexlab{}.
\newblock \showarticletitle{Technology-enhanced embodied learning}.
\newblock \bibinfo{journal}{\emph{Educational Technology \& Society}} \bibinfo{volume}{23}, \bibinfo{number}{3} (\bibinfo{year}{2020}), \bibinfo{pages}{81--94}.
\newblock


\bibitem[Johal et~al\mbox{.}(2020)]%
        {cellulo-sym}
\bibfield{author}{\bibinfo{person}{Wafa Johal}, \bibinfo{person}{Sonia Andersen}, \bibinfo{person}{Morgane Chevalier}, \bibinfo{person}{Ayberk Ozgur}, \bibinfo{person}{Francesco Mondada}, {and} \bibinfo{person}{Pierre Dillenbourg}.} \bibinfo{year}{2020}\natexlab{}.
\newblock \showarticletitle{Learning Symmetry with Tangible Robots}. In \bibinfo{booktitle}{\emph{Robotics in Education}}, \bibfield{editor}{\bibinfo{person}{Munir Merdan}, \bibinfo{person}{Wilfried Lepuschitz}, \bibinfo{person}{Gottfried Koppensteiner}, \bibinfo{person}{Richard Balogh}, {and} \bibinfo{person}{David Obdr{\v{z}}{\'a}lek}} (Eds.). \bibinfo{publisher}{Springer International Publishing}, \bibinfo{address}{Cham}, \bibinfo{pages}{270--283}.
\newblock
\showISBNx{978-3-030-26945-6}


\bibitem[Khodr et~al\mbox{.}(2020)]%
        {khodr2020allohaptic}
\bibfield{author}{\bibinfo{person}{Hala Khodr}, \bibinfo{person}{Soheil Kianzad}, \bibinfo{person}{Wafa Johal}, \bibinfo{person}{Aditi Kothiyal}, \bibinfo{person}{Barbara Bruno}, {and} \bibinfo{person}{Pierre Dillenbourg}.} \bibinfo{year}{2020}\natexlab{}.
\newblock \showarticletitle{AlloHaptic: Robot-Mediated Haptic Collaboration for Learning Linear Functions}. In \bibinfo{booktitle}{\emph{2020 29th IEEE International Conference on Robot and Human Interactive Communication (RO-MAN)}}. IEEE, \bibinfo{pages}{27--34}.
\newblock


\bibitem[Knight(1995)]%
        {knight1995vector}
\bibfield{author}{\bibinfo{person}{Randall~D Knight}.} \bibinfo{year}{1995}\natexlab{}.
\newblock \showarticletitle{The vector knowledge of beginning physics students}.
\newblock \bibinfo{journal}{\emph{The Physics Teacher}} \bibinfo{volume}{33}, \bibinfo{number}{2} (\bibinfo{year}{1995}), \bibinfo{pages}{74--77}.
\newblock


\bibitem[{McNeil} and Fyfe(2012)]%
        {mcneil_concreteness_2012}
\bibfield{author}{\bibinfo{person}{Nicole~M. {McNeil}} {and} \bibinfo{person}{Emily~R. Fyfe}.} \bibinfo{year}{2012}\natexlab{}.
\newblock \showarticletitle{Concreteness fading promotes transfer of mathematical knowledge}.
\newblock  \bibinfo{volume}{22}, \bibinfo{number}{6} (\bibinfo{year}{2012}), \bibinfo{pages}{440--448}.
\newblock
\showISSN{0959-4752}
\urldef\tempurl%
\url{https://doi.org/10.1016/j.learninstruc.2012.05.001}
\showDOI{\tempurl}


\bibitem[Nasir et~al\mbox{.}(2019)]%
        {Nasir:270077}
\bibfield{author}{\bibinfo{person}{Jauwairia Nasir}, \bibinfo{person}{Utku Norman}, \bibinfo{person}{Wafa Johal}, \bibinfo{person}{Jennifer~Kaitlyn Olsen}, \bibinfo{person}{Sina Shahmoradi}, {and} \bibinfo{person}{Pierre Dillenbourg}.} \bibinfo{year}{2019}\natexlab{}.
\newblock \showarticletitle{Robot Analytics: What Do Human-Robot Interaction Traces Tell Us About Learning?}
\newblock \bibinfo{journal}{\emph{[Proceedings of the IEEE RoMan 2019 - The 28th IEEE International Conference on Robot \&; Human Interactive Communication]}} (\bibinfo{year}{2019}).
\newblock
\urldef\tempurl%
\url{http://infoscience.epfl.ch/record/270077}
\showURL{%
\tempurl}


\bibitem[Neto et~al\mbox{.}(2020)]%
        {neto2020}
\bibfield{author}{\bibinfo{person}{Isabel Neto}, \bibinfo{person}{Wafa Johal}, \bibinfo{person}{Marta Couto}, \bibinfo{person}{Hugo Nicolau}, \bibinfo{person}{Ana Paiva}, {and} \bibinfo{person}{Arzu Guneysu}.} \bibinfo{year}{2020}\natexlab{}.
\newblock \showarticletitle{Using Tabletop Robots to Promote Inclusive Classroom Experiences}. In \bibinfo{booktitle}{\emph{Proceedings of the Interaction Design and Children Conference}} (London, United Kingdom) \emph{(\bibinfo{series}{IDC ’20})}. \bibinfo{publisher}{Association for Computing Machinery}, \bibinfo{address}{New York, NY, USA}, \bibinfo{pages}{281–292}.
\newblock
\showISBNx{9781450379816}
\urldef\tempurl%
\url{https://doi.org/10.1145/3392063.3394439}
\showDOI{\tempurl}


\bibitem[Nguyen and Meltzer(2003)]%
        {nguyen2003initial}
\bibfield{author}{\bibinfo{person}{Ngoc-Loan Nguyen} {and} \bibinfo{person}{David~E Meltzer}.} \bibinfo{year}{2003}\natexlab{}.
\newblock \showarticletitle{Initial understanding of vector concepts among students in introductory physics courses}.
\newblock \bibinfo{journal}{\emph{American journal of physics}} \bibinfo{volume}{71}, \bibinfo{number}{6} (\bibinfo{year}{2003}), \bibinfo{pages}{630--638}.
\newblock


\bibitem[Novack et~al\mbox{.}(2014)]%
        {novack2014action}
\bibfield{author}{\bibinfo{person}{Miriam~A Novack}, \bibinfo{person}{Eliza~L Congdon}, \bibinfo{person}{Naureen Hemani-Lopez}, {and} \bibinfo{person}{Susan Goldin-Meadow}.} \bibinfo{year}{2014}\natexlab{}.
\newblock \showarticletitle{From action to abstraction: Using the hands to learn math}.
\newblock \bibinfo{journal}{\emph{Psychological Science}} \bibinfo{volume}{25}, \bibinfo{number}{4} (\bibinfo{year}{2014}), \bibinfo{pages}{903--910}.
\newblock


\bibitem[{Oskiper} et~al\mbox{.}(2012)]%
        {Oskiper2012}
\bibfield{author}{\bibinfo{person}{T. {Oskiper}}, \bibinfo{person}{S. {Samarasekera}}, {and} \bibinfo{person}{R. {Kumar}}.} \bibinfo{year}{2012}\natexlab{}.
\newblock \showarticletitle{Multi-sensor navigation algorithm using monocular camera, IMU and GPS for large scale augmented reality}. In \bibinfo{booktitle}{\emph{2012 IEEE International Symposium on Mixed and Augmented Reality (ISMAR)}}. \bibinfo{pages}{71--80}.
\newblock


\bibitem[Ottmar and Landy(2017)]%
        {ottmar2017concreteness}
\bibfield{author}{\bibinfo{person}{Erin Ottmar} {and} \bibinfo{person}{David Landy}.} \bibinfo{year}{2017}\natexlab{}.
\newblock \showarticletitle{Concreteness fading of algebraic instruction: Effects on learning}.
\newblock \bibinfo{journal}{\emph{Journal of the Learning Sciences}} \bibinfo{volume}{26}, \bibinfo{number}{1} (\bibinfo{year}{2017}), \bibinfo{pages}{51--78}.
\newblock


\bibitem[{\"O}zg{\"u}r et~al\mbox{.}(2018)]%
        {ozgur2018declarative}
\bibfield{author}{\bibinfo{person}{Ayberk {\"O}zg{\"u}r}, \bibinfo{person}{Wafa Johal}, \bibinfo{person}{Arzu G{\"u}neysu~{\"O}zg{\"u}r}, \bibinfo{person}{Francesco Mondada}, {and} \bibinfo{person}{Pierre Dillenbourg}.} \bibinfo{year}{2018}\natexlab{}.
\newblock \showarticletitle{Declarative Physicomimetics for Tangible Swarm Application Development}. In \bibinfo{booktitle}{\emph{Proceedings of the 11th International Conference on Swarm Intelligence, ANTS 2018}}, Vol.~\bibinfo{volume}{11172}.
\newblock


\bibitem[{\"O}zg{\"u}r et~al\mbox{.}(2017a)]%
        {celluloTechIntro}
\bibfield{author}{\bibinfo{person}{Ayberk {\"O}zg{\"u}r}, \bibinfo{person}{Wafa Johal}, \bibinfo{person}{Francesco Mondada}, {and} \bibinfo{person}{Pierre Dillenbourg}.} \bibinfo{year}{2017}\natexlab{a}.
\newblock \showarticletitle{Haptic-enabled handheld mobile robots: Design and analysis}. In \bibinfo{booktitle}{\emph{Proceedings of the 2017 CHI Conference on Human Factors in Computing Systems}}. ACM, \bibinfo{pages}{2449--2461}.
\newblock


\bibitem[{\"O}zg{\"u}r et~al\mbox{.}(2017b)]%
        {ozgur2017cellulo}
\bibfield{author}{\bibinfo{person}{Ayberk {\"O}zg{\"u}r}, \bibinfo{person}{S{\'e}verin Lemaignan}, \bibinfo{person}{Wafa Johal}, \bibinfo{person}{Maria Beltran}, \bibinfo{person}{Manon Briod}, \bibinfo{person}{L{\'e}a Pereyre}, \bibinfo{person}{Francesco Mondada}, {and} \bibinfo{person}{Pierre Dillenbourg}.} \bibinfo{year}{2017}\natexlab{b}.
\newblock \showarticletitle{Cellulo: Versatile handheld robots for education}. In \bibinfo{booktitle}{\emph{Proceedings of the 2017 ACM/IEEE International Conference on Human-Robot Interaction}}. ACM, \bibinfo{pages}{119--127}.
\newblock


\bibitem[Piaget(1970)]%
        {piaget1970science}
\bibfield{author}{\bibinfo{person}{Jean Piaget}.} \bibinfo{year}{1970}\natexlab{}.
\newblock \showarticletitle{Science of education and the psychology of the child. Trans. D. Coltman.}
\newblock  (\bibinfo{year}{1970}).
\newblock


\bibitem[Pintaric and Kaufmann(2007)]%
        {pintaric2007affordable}
\bibfield{author}{\bibinfo{person}{Thomas Pintaric} {and} \bibinfo{person}{Hannes Kaufmann}.} \bibinfo{year}{2007}\natexlab{}.
\newblock \showarticletitle{Affordable infrared-optical pose-tracking for virtual and augmented reality}. In \bibinfo{booktitle}{\emph{Proceedings of Trends and Issues in Tracking for Virtual Environments Workshop, IEEE VR}}. \bibinfo{pages}{44--51}.
\newblock


\bibitem[Spiro(1988)]%
        {spiro1988multiple}
\bibfield{author}{\bibinfo{person}{Rand~J Spiro}.} \bibinfo{year}{1988}\natexlab{}.
\newblock \showarticletitle{Multiple analogies for complex concepts: Antidotes for analogy-induced misconception in advanced knowledge acquisition}.
\newblock \bibinfo{journal}{\emph{Center for the Study of Reading Technical Report; no. 439}} (\bibinfo{year}{1988}).
\newblock


\bibitem[Suh et~al\mbox{.}(2020)]%
        {suh2020we}
\bibfield{author}{\bibinfo{person}{Sangho Suh}, \bibinfo{person}{Martinet Lee}, {and} \bibinfo{person}{Edith Law}.} \bibinfo{year}{2020}\natexlab{}.
\newblock \showarticletitle{How do we design for concreteness fading? survey, general framework, and design dimensions}. In \bibinfo{booktitle}{\emph{Proceedings of the Interaction Design and Children Conference}}. \bibinfo{pages}{581--588}.
\newblock


\bibitem[Tatarian et~al\mbox{.}(2020)]%
        {tatarian2020mobiaxis}
\bibfield{author}{\bibinfo{person}{Karen Tatarian}, \bibinfo{person}{Sebastian Wallkotter}, \bibinfo{person}{Sera Buyukgoz}, \bibinfo{person}{Rebecca Stower}, {and} \bibinfo{person}{Mohamed Chetouani}.} \bibinfo{year}{2020}\natexlab{}.
\newblock \showarticletitle{MobiAxis: An Embodied Learning Task for Teaching Multiplication with a Social Robot}.
\newblock \bibinfo{journal}{\emph{arXiv preprint arXiv:2004.07806}} (\bibinfo{year}{2020}).
\newblock


\bibitem[Trory et~al\mbox{.}(2018)]%
        {trory2018designing}
\bibfield{author}{\bibinfo{person}{Anthony Trory}, \bibinfo{person}{Kate Howland}, {and} \bibinfo{person}{Judith Good}.} \bibinfo{year}{2018}\natexlab{}.
\newblock \showarticletitle{Designing for concreteness fading in primary computing}. In \bibinfo{booktitle}{\emph{Proceedings of the 17th ACM Conference on Interaction Design and Children}}. ACM, \bibinfo{pages}{278--288}.
\newblock


\bibitem[Tsai and Hsu(2016)]%
        {tsai2016application}
\bibfield{author}{\bibinfo{person}{Chang-Yen Tsai} {and} \bibinfo{person}{Kuo-Hsun Hsu}.} \bibinfo{year}{2016}\natexlab{}.
\newblock \showarticletitle{An application of using bluetooth indoor positioning, image recognition and augmented reality}. In \bibinfo{booktitle}{\emph{2016 IEEE 13th International Conference on e-Business Engineering (ICEBE)}}. IEEE, \bibinfo{pages}{276--281}.
\newblock


\bibitem[Uchiyama and Saito(2011)]%
        {uchiyama2011random}
\bibfield{author}{\bibinfo{person}{Hideaki Uchiyama} {and} \bibinfo{person}{Hideo Saito}.} \bibinfo{year}{2011}\natexlab{}.
\newblock \showarticletitle{Random dot markers}. In \bibinfo{booktitle}{\emph{2011 IEEE Virtual Reality Conference}}. IEEE, \bibinfo{pages}{35--38}.
\newblock


\bibitem[Wikipedia(2021)]%
        {wiki:Mathematical_anxiety}
\bibfield{author}{\bibinfo{person}{Wikipedia}.} \bibinfo{year}{2021}\natexlab{}.
\newblock \bibinfo{title}{{Mathematical anxiety} --- {W}ikipedia{,} The Free Encyclopedia}.
\newblock \bibinfo{howpublished}{\url{http://en.wikipedia.org/w/index.php?title=Mathematical\%20anxiety&oldid=1050447757}}.
\newblock
\newblock
\shownote{[Online; accessed 22-October-2021]}.


\bibitem[Wutchana and Emarat(2011)]%
        {wutchana2011students}
\bibfield{author}{\bibinfo{person}{Umporn Wutchana} {and} \bibinfo{person}{Narumon Emarat}.} \bibinfo{year}{2011}\natexlab{}.
\newblock \showarticletitle{Students’ understanding of graphical vector addition in one and two dimensions}.
\newblock \bibinfo{journal}{\emph{Eurasian Journal of Physics and Chemistry Education}} \bibinfo{volume}{3}, \bibinfo{number}{2} (\bibinfo{year}{2011}), \bibinfo{pages}{102--111}.
\newblock


\bibitem[Xia and Zhong(2018)]%
        {xia2018systematic}
\bibfield{author}{\bibinfo{person}{Liying Xia} {and} \bibinfo{person}{Baichang Zhong}.} \bibinfo{year}{2018}\natexlab{}.
\newblock \showarticletitle{A systematic review on teaching and learning robotics content knowledge in K-12}.
\newblock \bibinfo{journal}{\emph{Computers \& Education}}  \bibinfo{volume}{127} (\bibinfo{year}{2018}), \bibinfo{pages}{267--282}.
\newblock


\end{thebibliography}

\newpage
\section*{Appendix}
\subsection{Tables}

\begin{minipage}{0.96\textwidth}
\begin{minipage}[b]{0.47\textwidth}
\centering
    \begin{tabular}[b]{llrrll}\toprule{} & device &  Age1 &  Age2 & Gender1 & Gender2 \\\midrule0  &   0b78 &   NaN &   NaN &    Male &    Male \\1  &   0b78 &  17.0 &  17.0 &  Female &  Female \\2  &   35eb &   NaN &   NaN &    Male &    Male \\3  &   35eb &  17.0 &  17.0 &    Male &   Other \\4  &   3693 &  14.0 &  15.0 &  Female &  Female \\5  &   3693 &  16.0 &   NaN &    Male &    Male \\6  &   e8c1 &  14.0 &  15.0 &   Other &  Female \\7  &   e8c1 &  16.0 &   NaN &    Male &    Male \\8  &   nv07 &  14.0 &  15.0 &    Male &    Male \\9  &   nv07 &  17.0 &  16.0 &    Male &  Female \\10 &   nv09 &  17.0 &  16.0 &  Female &  Female \\11 &   nv10 &  15.0 &  15.0 &    Male &    Male \\12 &   nv10 &  17.0 &  17.0 &    Male &    Male \\13 &   nv11 &  15.0 &  16.0 &    Male &    Male \\14 &   nv11 &  16.0 &  17.0 &    Male &    Male \\\bottomrule\end{tabular}
    \captionof{table}{Groups and participants' demographic information}
\label{tab:participants}
\end{minipage}
\end{minipage}

\subsection{Figures}

\begin{figure}[ht]
    \setlength{\belowcaptionskip}{50em}
    \centering
    \includegraphics[width=0.7\paperwidth]{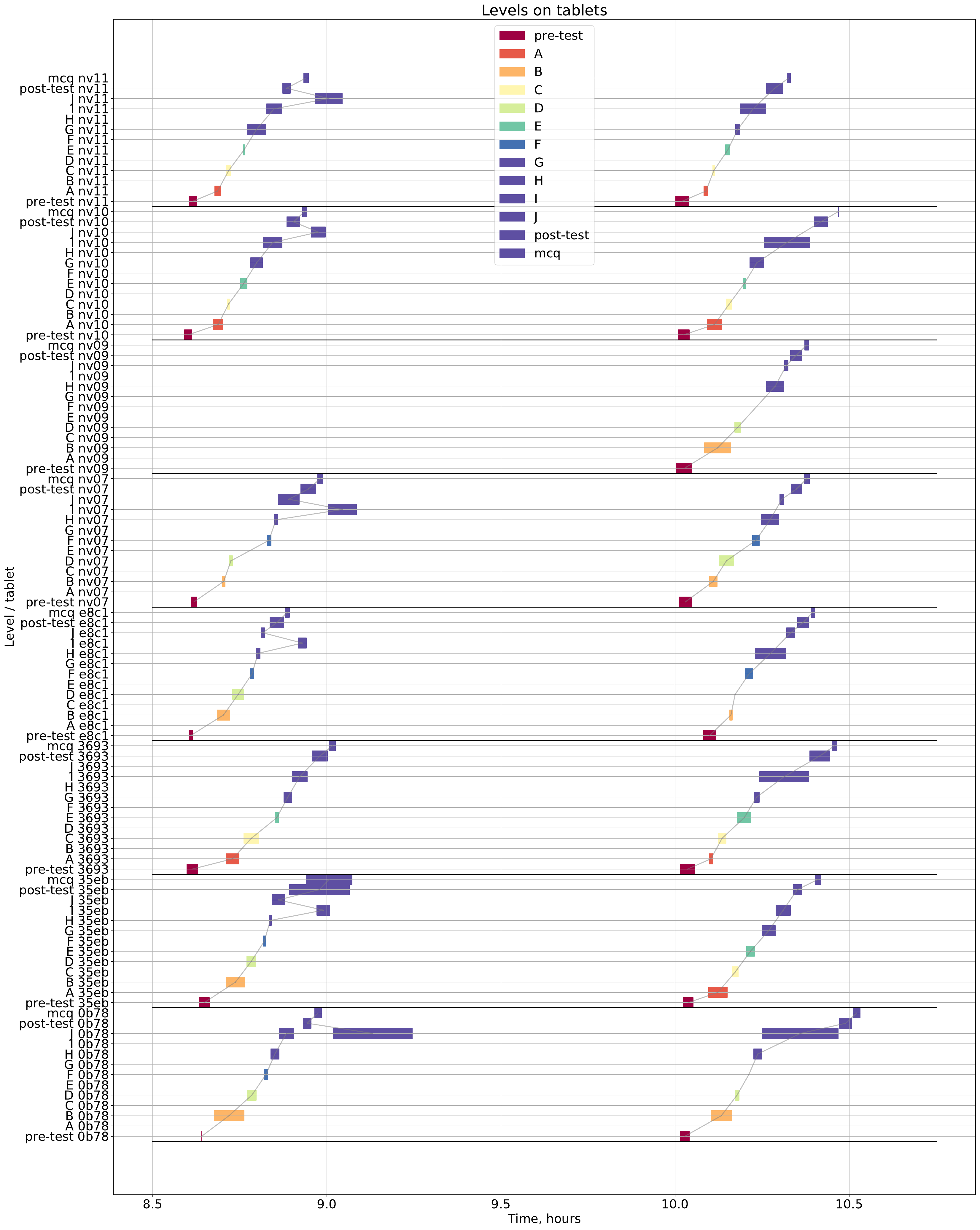}
    \caption{Timeline of the gameplay over devices for both groups}
    \label{fig:timeline}
\end{figure}

\begin{figure}[h!]
	\centering
\includegraphics[width=0.3\textwidth]{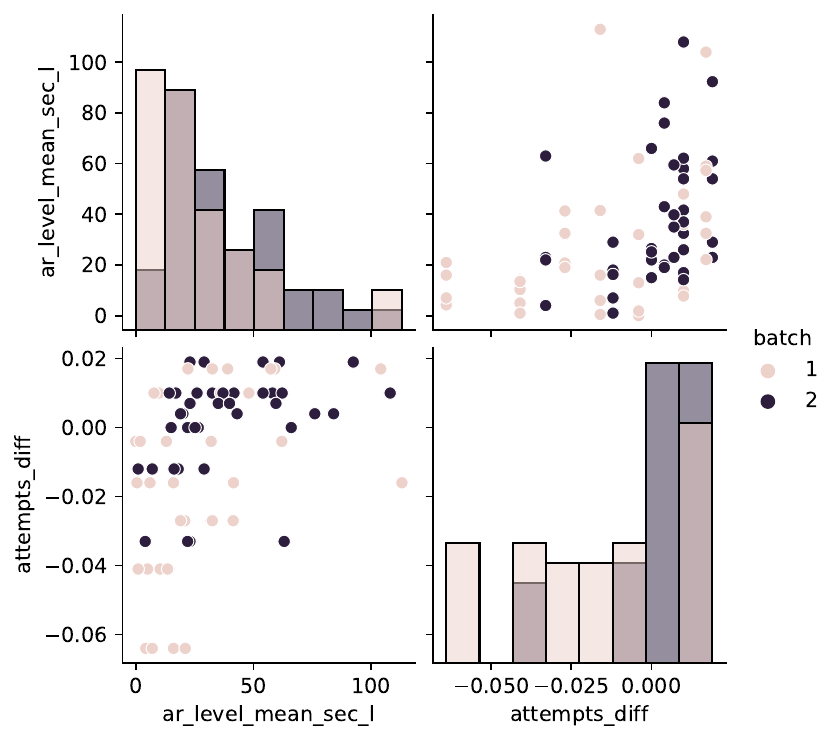} \hfill
\includegraphics[width=0.3\textwidth]{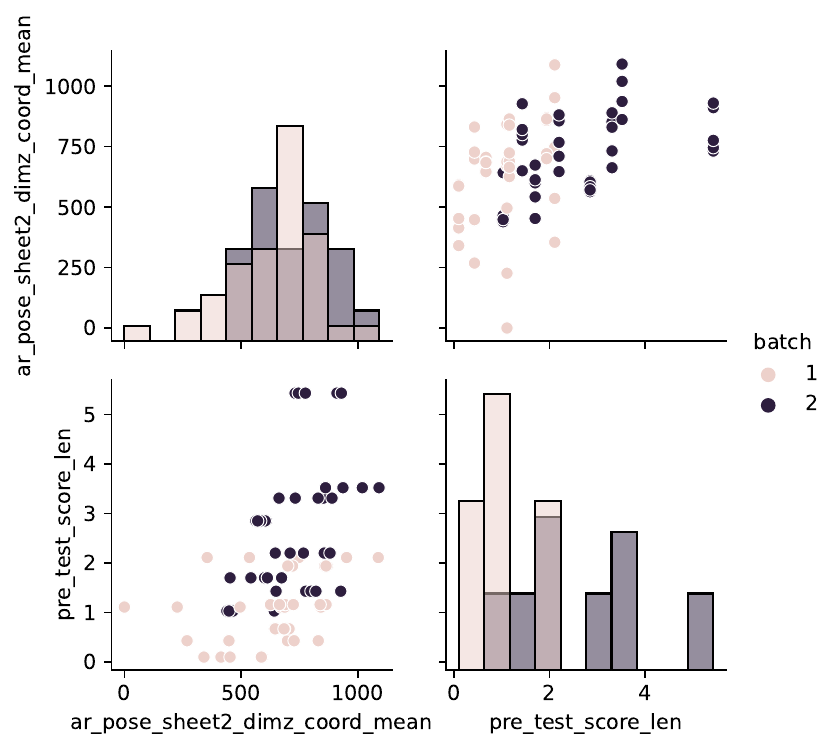}
\hfill
\includegraphics[width=0.3\textwidth]{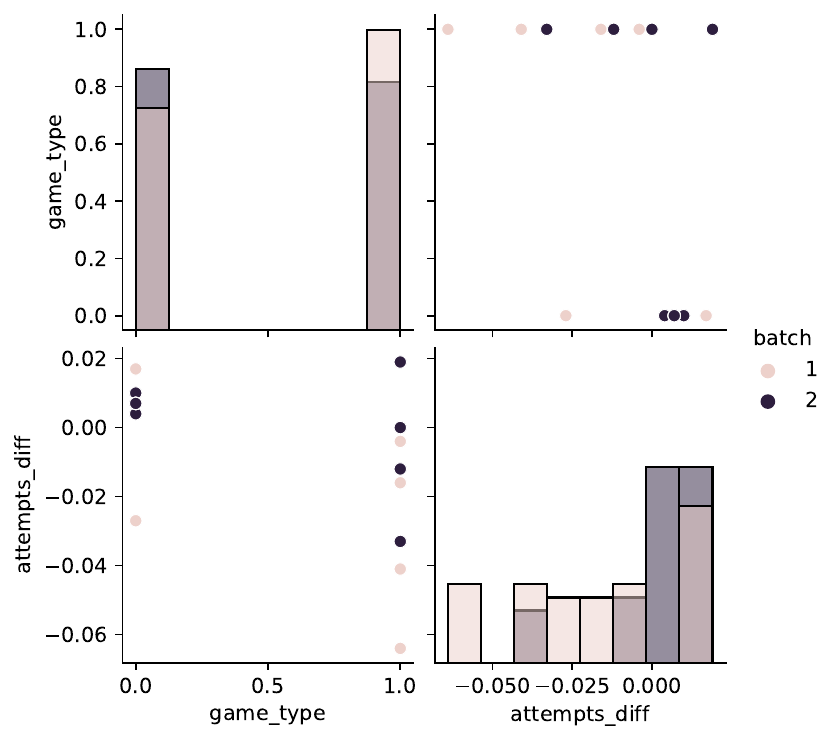}\\

\includegraphics[width=0.3\textwidth]{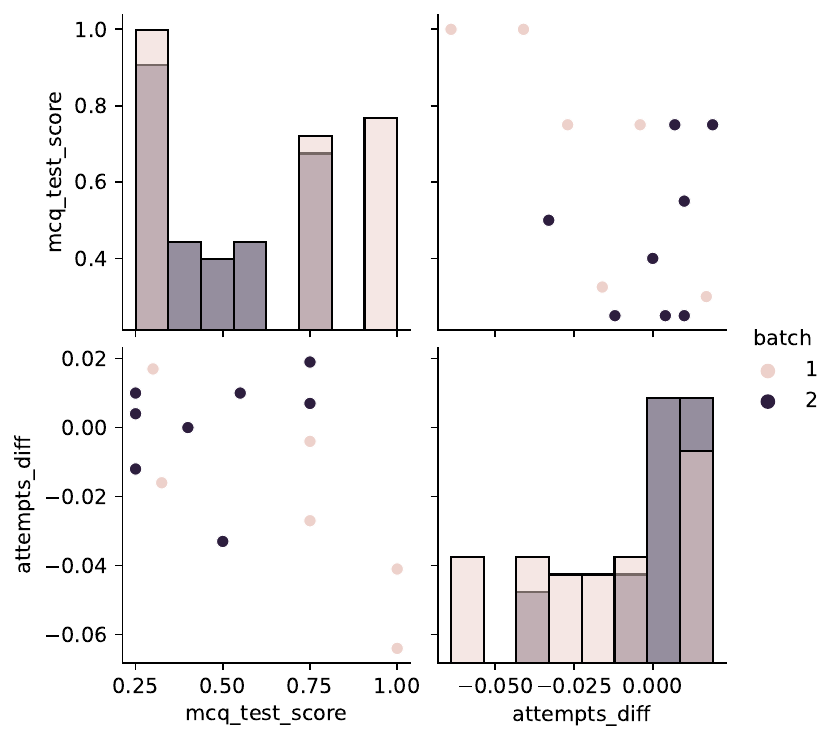}
\hfill
\includegraphics[width=0.3\textwidth]{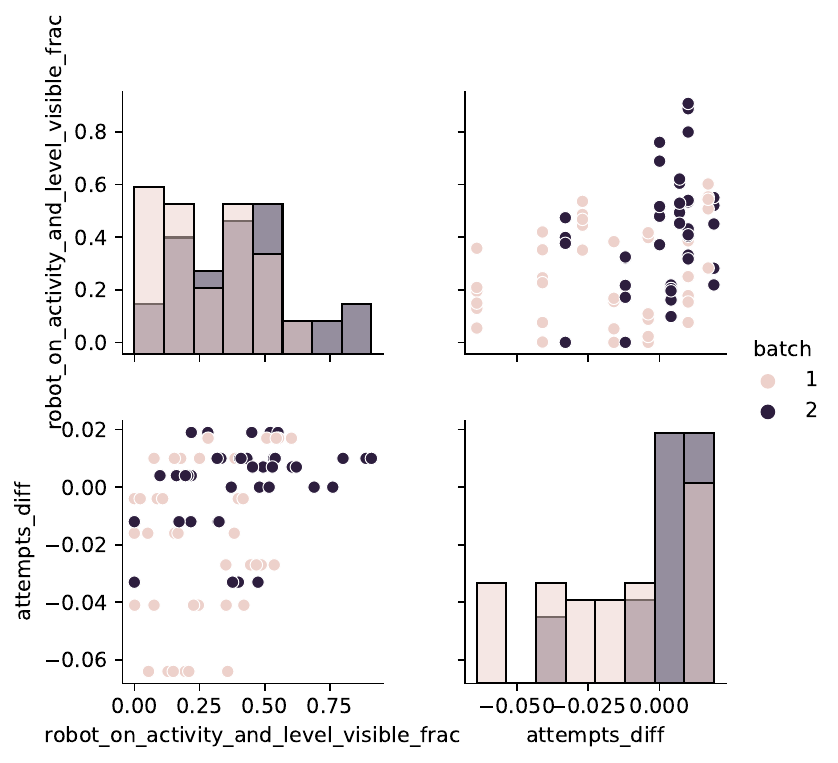}
\hfill
\includegraphics[width=0.3\textwidth]{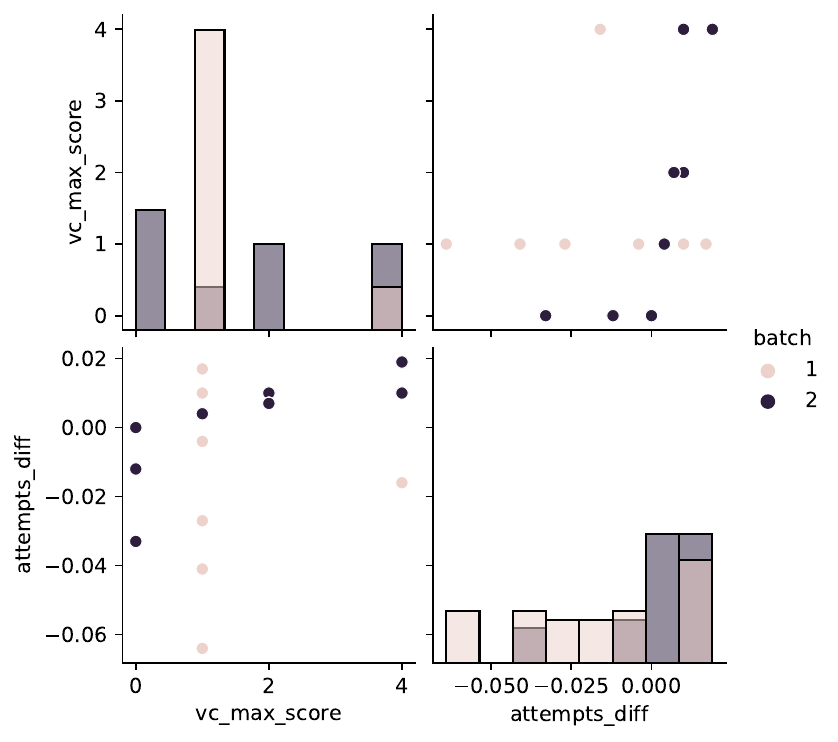}\\

\includegraphics[width=0.3\textwidth]{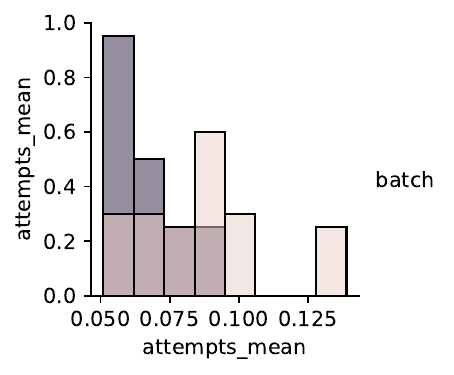}
\hfill
\includegraphics[width=0.3\textwidth]{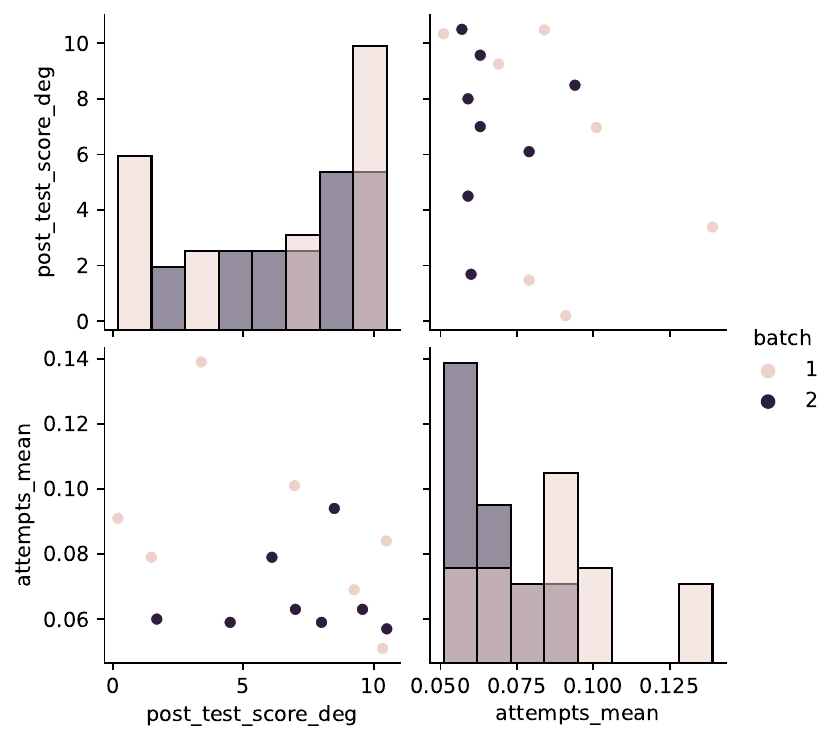}
\hfill
\includegraphics[width=0.3\textwidth]{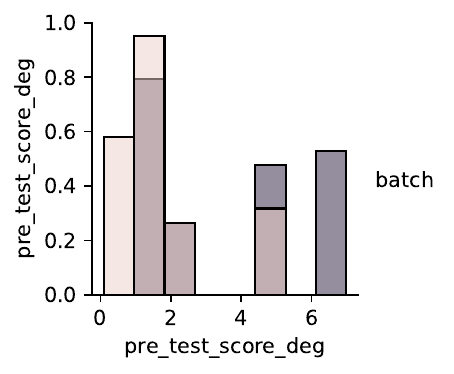}\\

\includegraphics[width=0.3\textwidth]{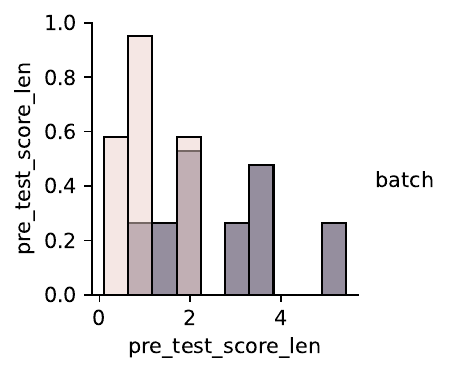}
\hfill
\includegraphics[width=0.3\textwidth]{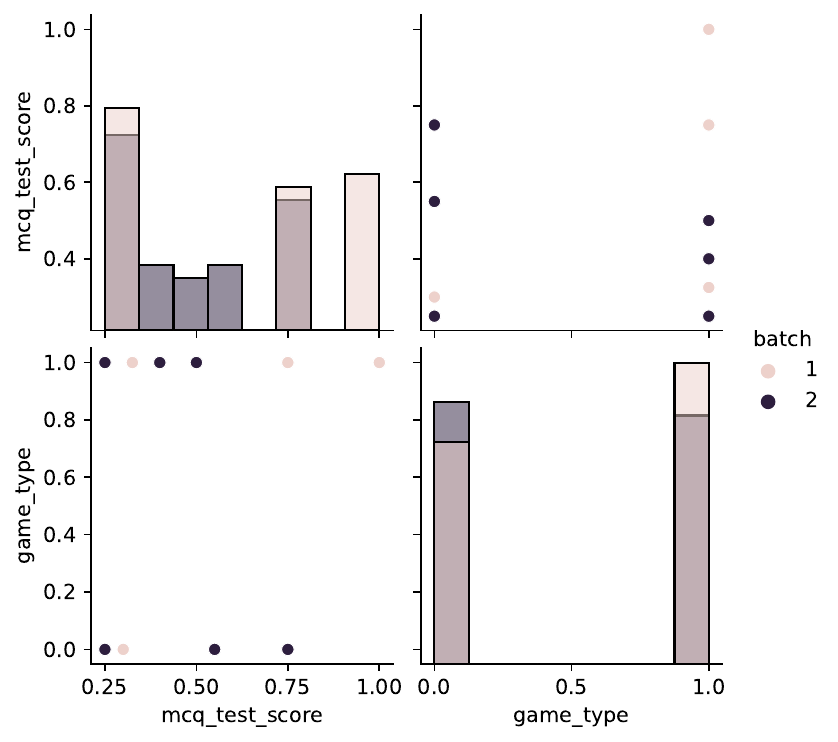}
\hfill
\includegraphics[width=0.3\textwidth]{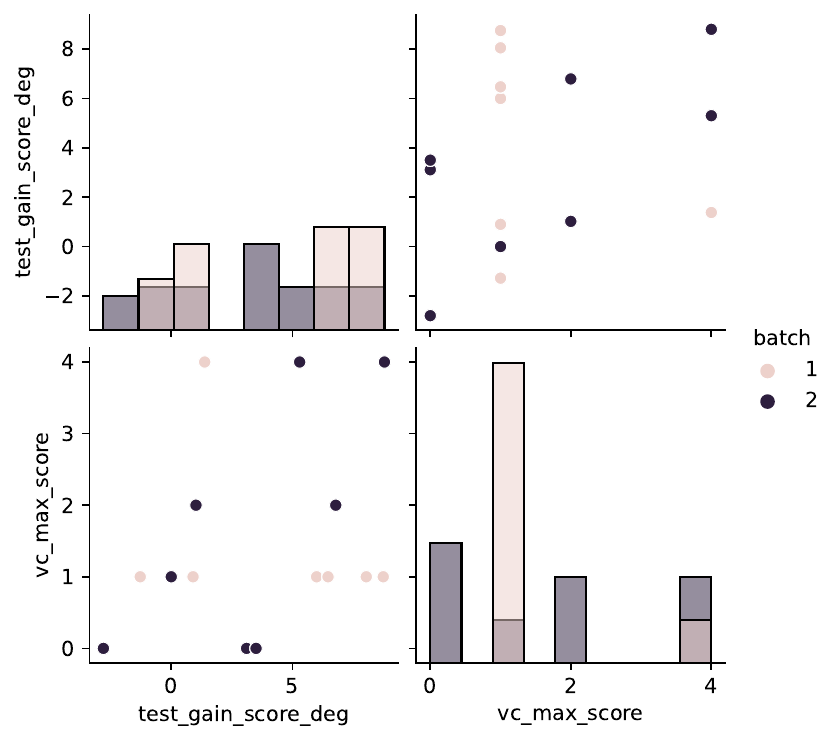}

	\caption{XY plots for the metrics used in the experiment}
	\label{fig:xy_plots}
\end{figure}

\begin{figure}[ht]
    \centering
\includegraphics[width=0.7\textwidth]{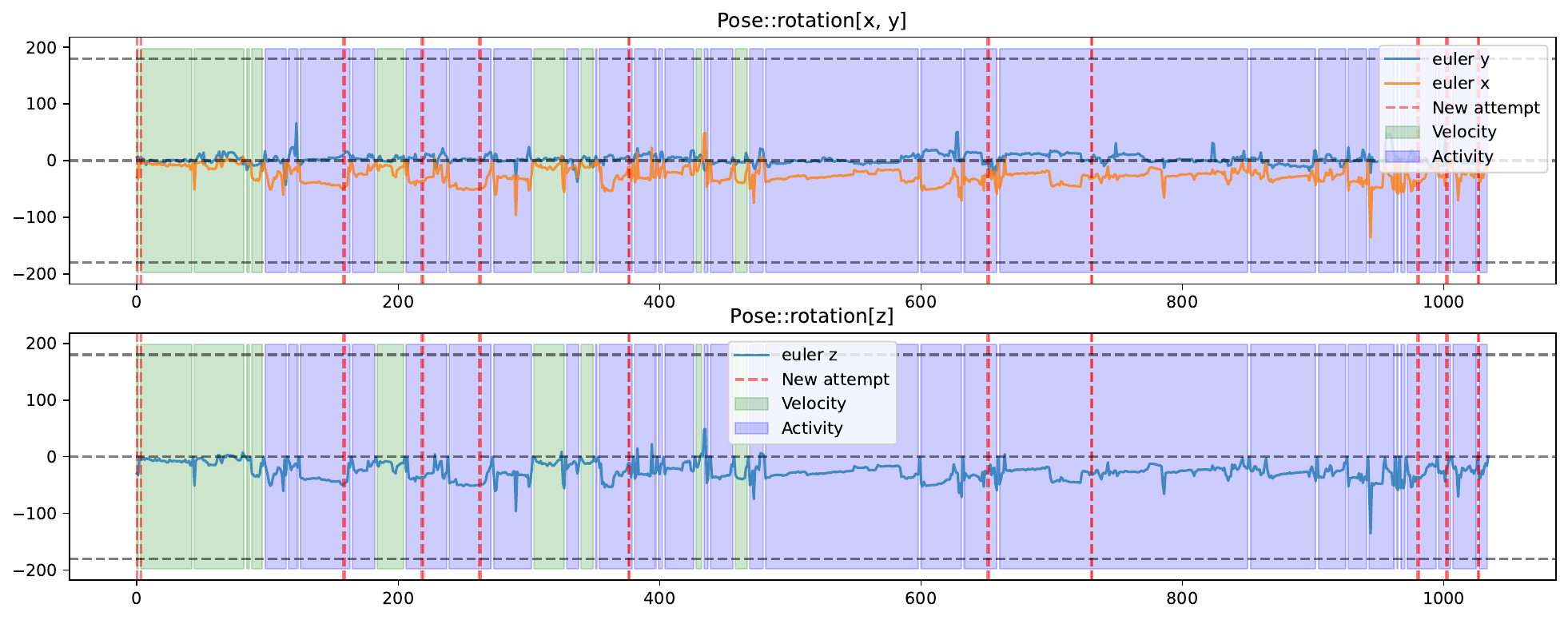}\\
\includegraphics[width=0.7\textwidth]{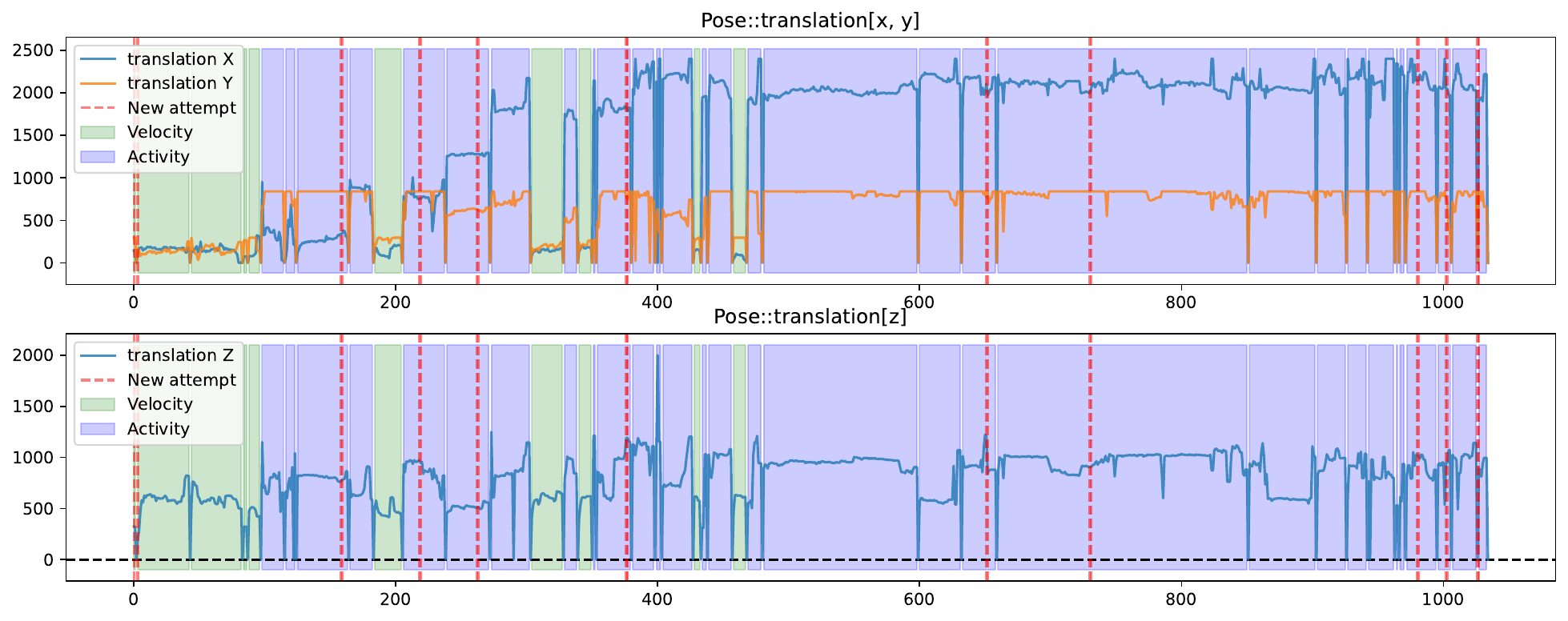}
\caption{Timelines for one of the games for AR metrics. Top two charts show rotation (Euler angles for $x$ and $y$ axes, first chart, and vertical $z$ axis, second chart). Last two charts show translation (with respect to the top left corner of the activity sheet), third chart shows the $x$ and $y$ axes, and the final chart shows the vertical $z$ axis. Red dashed lines indicate a new attempt, green regions indicate the velocity setting sheet visible, and the blue regions indicate the main activity sheet visible.}
\label{fig:ar_example_timeline}
\end{figure}

\begin{figure}[ht]
    \centering
    \includegraphics[width=0.95\textwidth]{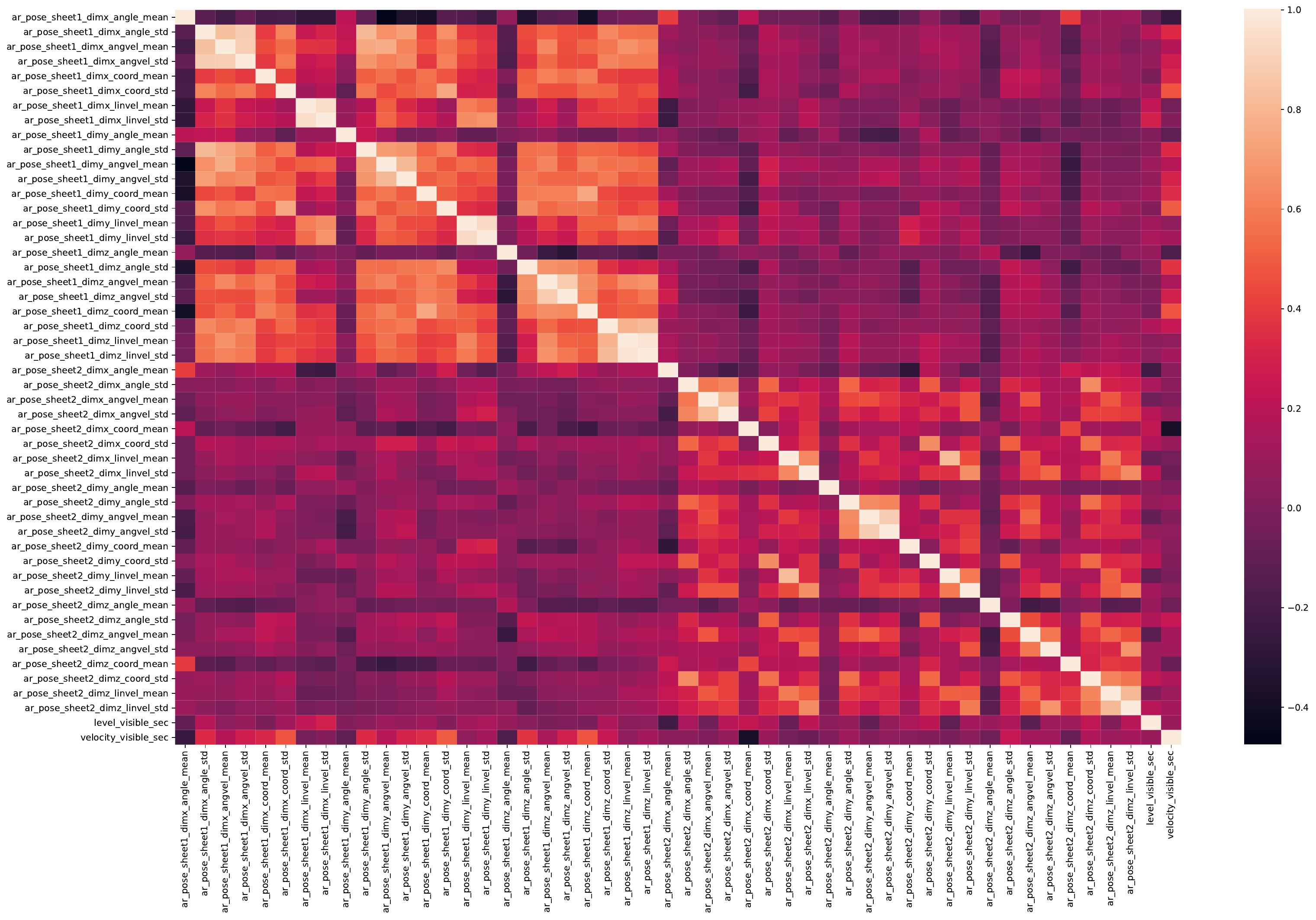}
    \caption{Correlation coefficients between all AR metrics (angle, coordinate, their derivatives). High values for sheet 1 and sheet 2, but not between them indicate that the tablet is used differently on these paper sheets.}
    \label{fig:ar_post_correlations}
\end{figure}

\begin{figure}
	\centering
	\includegraphics[width=0.5\textwidth]{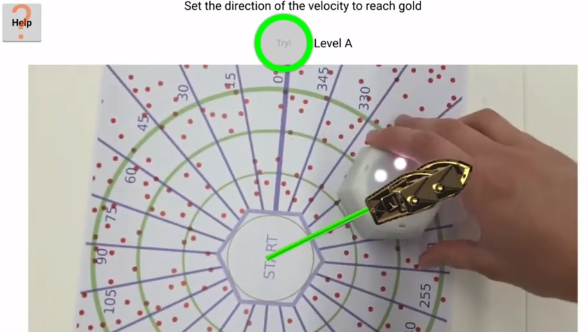}
	\caption{The velocity setting stage}
	\label{fig:game_velocity}
\end{figure}

\begin{figure}
	\centering
	\includegraphics[width=0.5\textwidth]{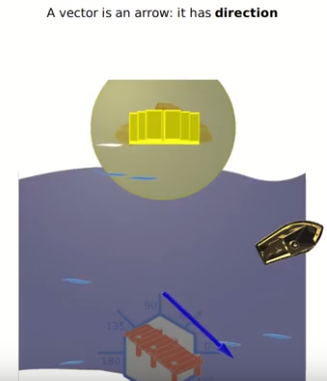}
	\caption{An interactive lecture explaining vectors}
	\label{fig:game_lecture}
\end{figure}

\begin{figure}
    \centering
    \includegraphics[width=0.22\textwidth]{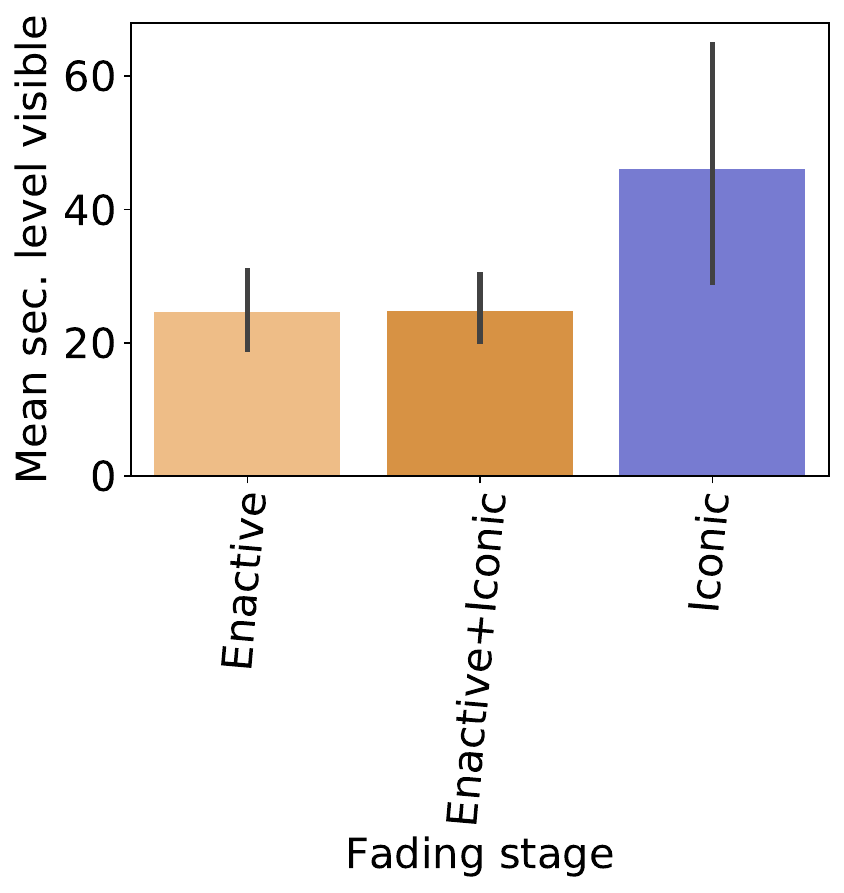}
    \includegraphics[width=0.22\textwidth]{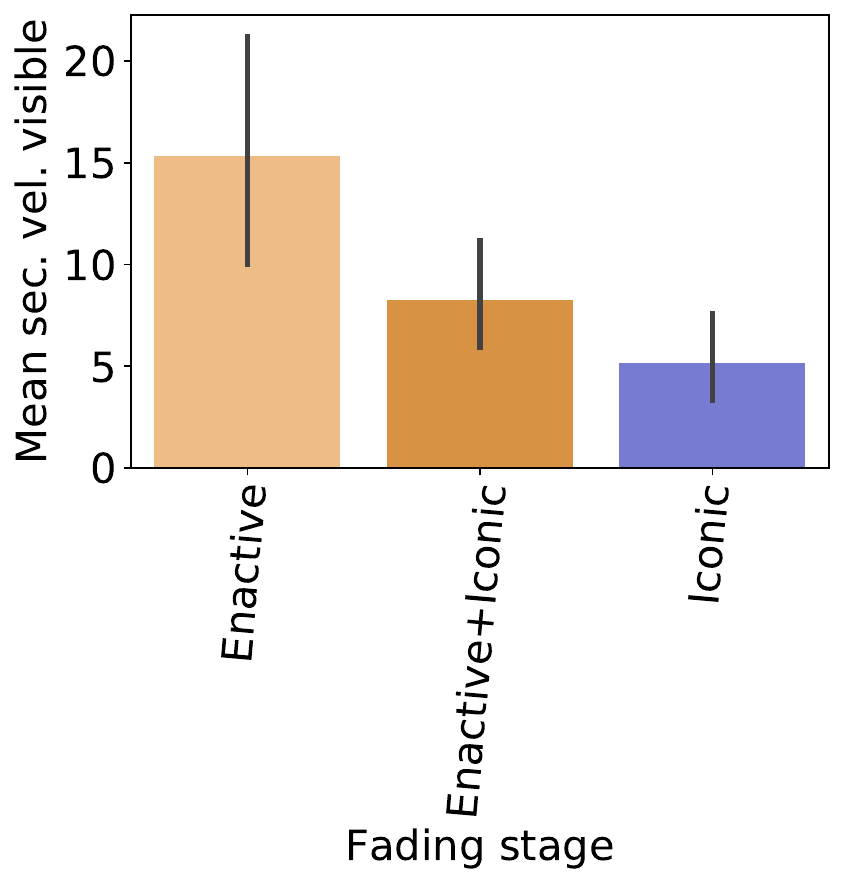}
    \includegraphics[width=0.22\textwidth]{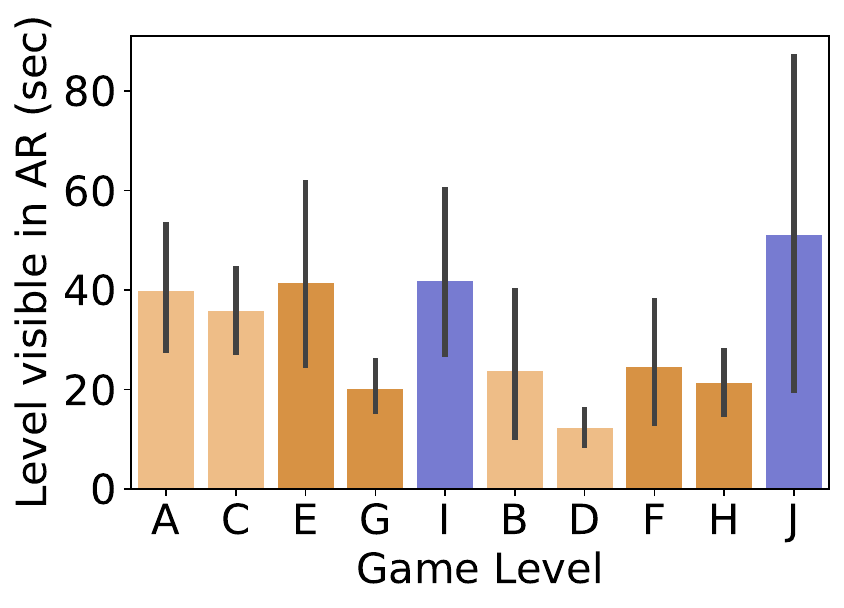}
    \includegraphics[width=0.22\textwidth]{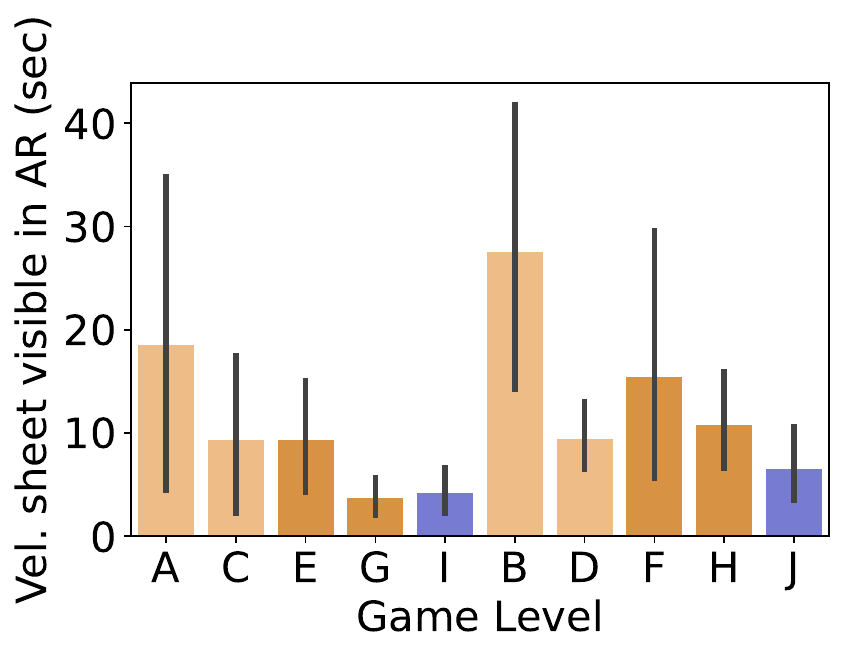}
    \caption{AR detailed timing data for velocity sheet and the level. On the x-axis are either level concreteness fading stages or the levels. There is no difference in time spent for levels for the first two stages, while it takes more time to solve the abstract stage (leftmost chart). Time to set the velocity decreases and we hypothesize that it represents learning how to use the system (second chart).}
    \label{fig:ar_usage}
\end{figure}
    \begin{figure}[ht]
         \centering
         \includegraphics[width=0.4\textwidth]{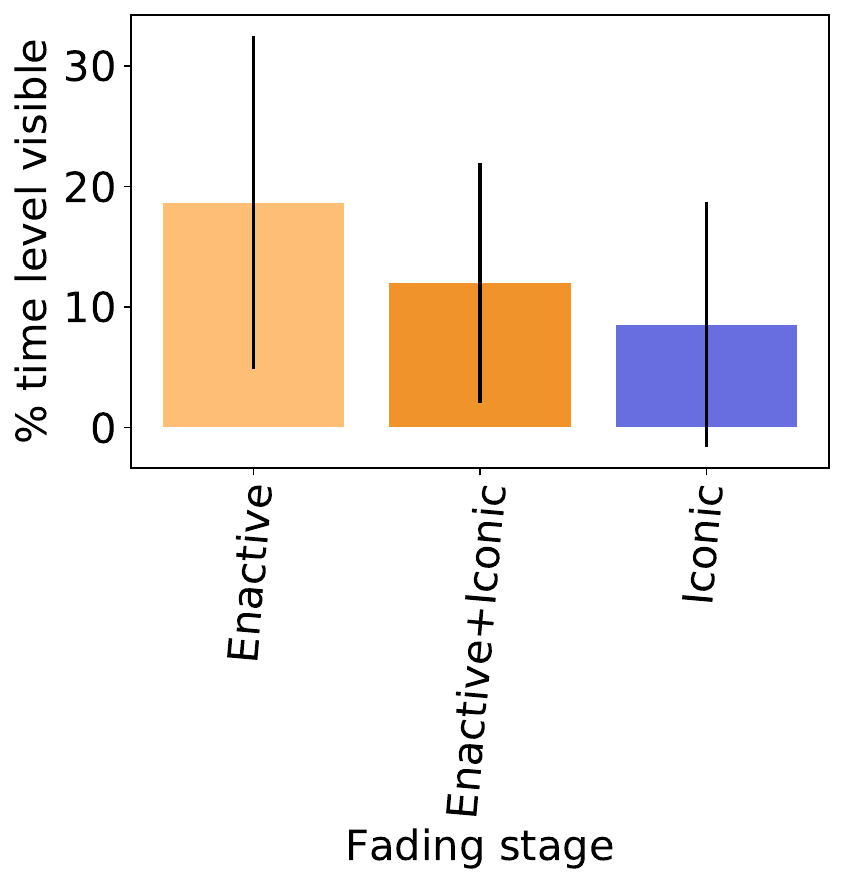}         
         \caption{Total time spent on each stage by teams in the AR mode (with the AR information visible on the tablet)}
         \label{fig:time_ar}
     \end{figure}
     \begin{figure}[t]
         \centering
         \includegraphics[width=0.4\textwidth]{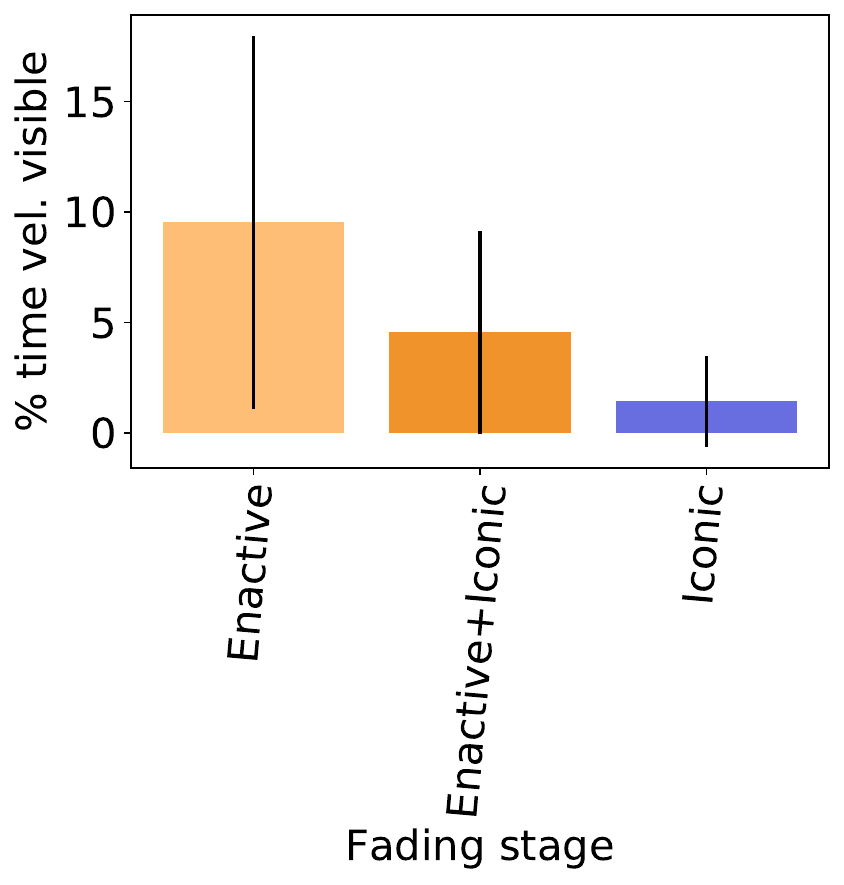}       
         \caption{Percentage of time spent on the velocity sheet in AR}
         \label{fig:time_ar_velocity}
     \end{figure}
     \hfill
     \begin{figure}[t]
         \centering
         \includegraphics[width=0.4\textwidth]{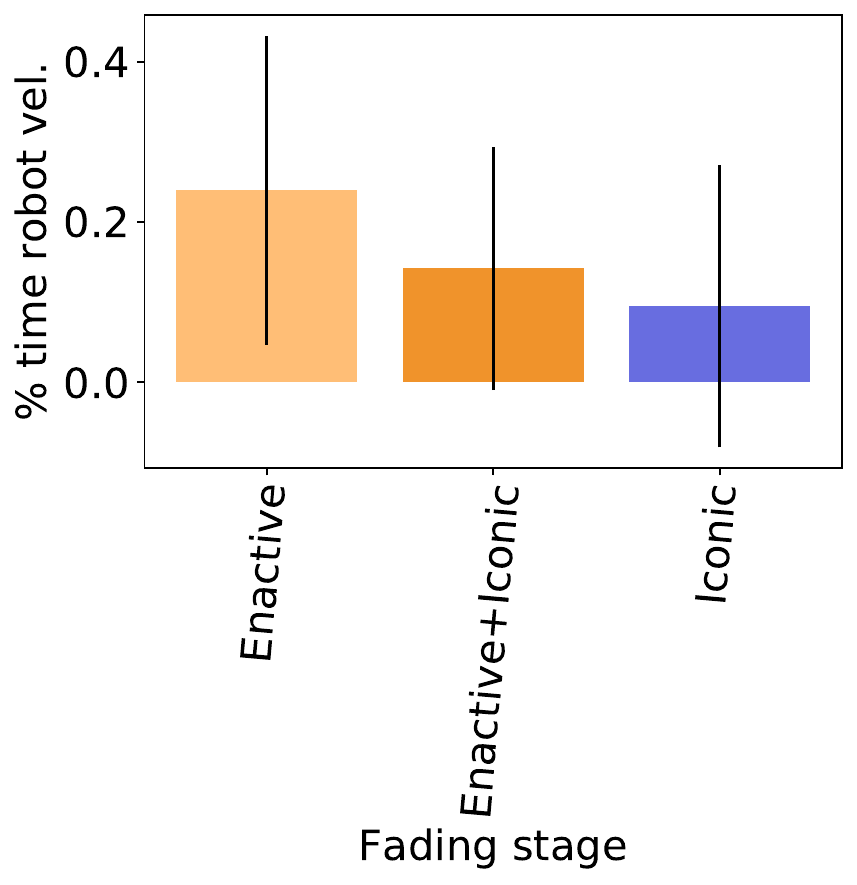}    
         \caption{Percentage of time spent on setting the velocity with the robot at each stage}
         \label{fig:time_ar_velocity_robot}
     \end{figure}
    \begin{figure}[t]
         \centering
           \includegraphics[width=0.4\textwidth]{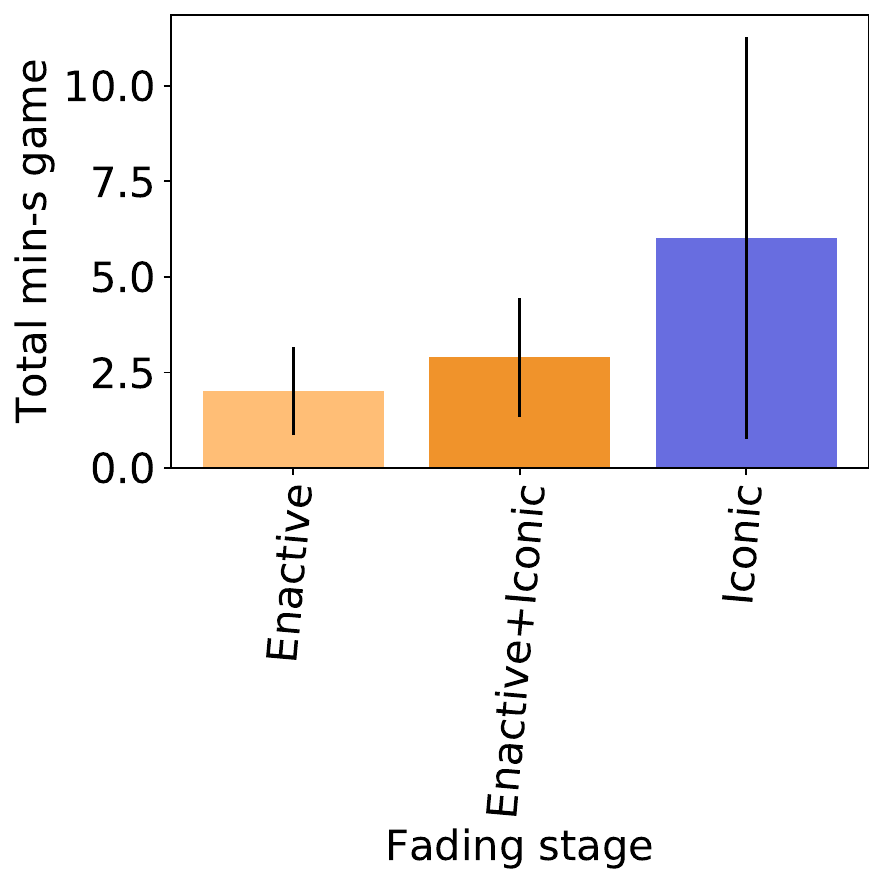}      
         \caption{Total time spent on each stage}
         \label{fig:time_stages}
     \end{figure}

\end{document}